\documentclass[hidelinks,journal]{IEEEtran}
\usepackage{url,array,lipsum,hhline,graphicx,subfigure,amsfonts,cite,epstopdf,makeidx,outlines,textcomp,multirow,soul,hyperref,enumitem,amssymb,amsmath,color,graphicx, amsbsy,}
\usepackage[table,xcdraw]{xcolor}
\usepackage[linesnumbered,ruled]{algorithm2e}
\usepackage[noprefix]{nomencl}
\usepackage[none]{hyphenat}
\allowdisplaybreaks

\begin{document}
\bstctlcite{IEEEexample:BSTcontrol}

\title{Coordinating Flexible Ramping Products with Dynamics of the Natural Gas Network}

\author{ {Reza Bayani, \emph{Student Member, IEEE} and Saeed D. Manshadi, \emph{Member,  IEEE}}
\thanks{Reza Bayani is with University of California San Diego, La Jolla, CA, 92093, USA, and San Diego State University. Saeed D. Manshadi is with San Diego State University, San Diego, CA, 92182, USA. e-mail: rbayani@ucsd.edu; smanshadi@sdsu.edu.
\vspace{-0.35cm}
}
}
\markboth{}
{Shell \MakeLowercase{\textit{et al.}}: Bare Demo of IEEEtran.cls for Journals}

\maketitle

\begin{abstract}

In electricity networks with high penetration levels of renewable resources, Flexible Ramping Products (FRPs) are among the utilized measures for dealing with the potential fluctuations in the net demand. This paper investigates the impacts of FRPs on the operation of interdependent electricity and natural gas networks. To accurately model and reflect the effects of variations in the natural gas fuel demand on the natural gas network, a dynamic Optimal Gas Flow (OGF) formulation is utilized.
The non-convex dynamic model of the natural gas system is represented in a convex form via a tight relaxation scheme. An improved distributed optimization
method is proposed to solve the coordinated operation problem in a privacy-preserving manner, where the two infrastructures only share limited information.
We introduce the Inexact Varying Alternating Direction Method of Multipliers (IV-ADMM) and show that compared with the classic ADMM, it converges considerably faster and in fewer iterations.
Through a comparison of day-ahead and real-time operation planning results, it is concluded that without accounting for natural gas network dynamics, the FRP model is not a trustworthy tool in day-ahead planning against uncertainties. 
\end{abstract}
\vspace{-0.15cm}
\IEEEpeerreviewmaketitle

\section*{Nomenclature}
\vspace{-0.15cm}
\subsection*{Indices, Sets, and Superscripts}
\noindent \begin{tabular}{ll}
$b,\mathcal{B}$ & Buses in the  electricity network\\
$c,\mathcal{C}$ & Compressors in the natural gas network\\
$g,\mathcal{G},G$ & Generation units in the electricity network\\
$h,\mathcal{H},H$ & Heat demand in the natural gas network\\
$j,\mathcal{J},J$ & Junctions in the natural gas network\\
$l,\mathcal{L},L$ & Lines in the electricity network\\
$m,\mathcal{M}$ & Pipe segments in the natural gas network\\
$p,\mathcal{P}$ & Pipes in the natural gas network\\
$s,\mathcal{S},S$ & Natural gas suppliers \\
$t,\mathcal{T}$ & Time steps\\
$u,\mathcal{U},U$ & Gas-fired units\\
$v,\mathcal{V},V$ & Solar photovoltaic units\\
$w,\mathcal{W},W$ & Wind turbine units
\end{tabular}
\vspace{-0.15cm}
\subsection*{Variables}
\noindent \begin{tabular}{ll}
$d$ & Natural gas demand in the natural gas network \\
$f$ & Mass flow rate of natural gas\\
$p$ & Dispatched electricity power of units/ loads\\
$q$ & Natural gas supplier output\\
$r$ & Flexible ramping product of gas-fired units\\
$\theta$ & Voltage angle of electricity bus\\
$\pi$ & Pressure of natural gas within pipelines/ junctions\\
$\tau,\sigma,\eta,\zeta$ & Dual variables
\end{tabular}

\vspace{1cm}
\section{Introduction} \label{sec:intro} 
\IEEEPARstart{O}NE of the inherent characteristics of renewable energy sources (RESs) is their non-dispatchable volatile power generation. In a RES integrated electricity network, net demand is defined as the total demand minus the total generation of non-dispatchable units. With the rise in the penetration level of RESs in an electricity network, the variations in the net demand will increase. Thus, it is necessary to leverage dispatchable generation units to mitigate the challenges introduced by the extensive RES penetration \cite{navid2012market}. Failure to follow the net demand imposes undesirable economical and operational consequences on the system operator such as fluctuations in market price and load shedding \cite{wang2014flexible}. In various research works, Flexible Ramping Products (FRPs) are investigated as an effective means of meeting net demand fluctuations in a RES integrated system. The successful implementation of the flexible ramp constraint in these studies suggests that a mix of operational, market, and policy solutions are required to provide systemic ability to meet uncertainties in RES generation.
\vspace{-0.4cm}
\subsection{\textbf{Background and Motivation}} \vspace{-0.1cm}
Natural gas-fired generation units are practically the main source of electricity network flexibility due to their intrinsic fast ramping up/down rates. According to the Energy Information Administration, natural gas is the prevalent source of electric power generation in the United States. The natural gas-based electricity generation is projected to steadily account for more than a third of the total power generation in the United States over the next 30 years \cite{eia2021}. Based on this report, the share of RESs is expected to double by 2050 (i.e. from the currently 21\% share of electricity generation to 42\%), replacing natural gas as the leader of electrical power source types. The current work is motivated by these reports, which highlight the growing importance of FRPs as a measure to facilitate renewable integration in an electricity network dominated by gas-fired units.

\par Some independent system operators such as CAISO have added FRPs to their markets to enhance generation flexibility \cite{wang2015real}. The benefits of utilizing thermal units as sources of ramping services, in combination with energy storage systems and demand response in providing flexible generation are studied in \cite{wu2014thermal}. 
Electric vehicles \cite{zhang2015impact}, spinning reserve services\cite{khoshjahan2019harnessing}, and bulk storage systems \cite{heydarian2017optimal} are also investigated as means of offering FRPs in the literature. Other aspects of FRPs such as transmission expansion planning considering ramping requirements \cite{li2017robust}, ramping resources in day-ahead market \cite{khodayar2016multiple}, and non-deterministic flexible ramp reserves\cite{alizadeh2017multistage} are studied in the literature as well.

\vspace{-0.45cm}
\subsection{\textbf{The Gap in the Existing Literature}}
\vspace{-0.1cm} From the flow equation point of view, the Optimal Gas Flow (OGF) problem utilized in the existing literature of coordinated operation problem is best categorized as \textit{steady-state} and \textit{dynamic} models. The coordinated operation problem of interconnected electricity and natural gas networks has been extensively investigated such as security-constrained approach \cite{correa2014security}, high renewable penetration case \cite{badakhshan2019impact}, and a convex relaxation method \cite{manshadi2018tight}. However, these works consider natural gas steady-state model, which \emph{fails to properly model how the natural gas network dynamics respond to changes in the demand side} \cite{bayani2022natural}.
Unlike the electricity network, \emph{the natural gas network does \ul{not} serve the variations in demand with instantaneous adjustments in supply}. In a natural gas network that supplies the fuel demand of gas-fired units, the amount of gas withdrawn from pipelines is subject to variations. Consequently, it is crucial to utilize accurate models for the dynamics of natural gas within pipelines and figure out how changes in demand are served through the gas stored in the pipelines. 

\par Coordination of the electricity and natural gas networks with dynamic OGF model is investigated thoroughly with applications such as microgrids \cite{xu2014dynamic}, bi-directional energy conversion \cite{fang2017dynamic}, robust generation scheduling \cite{yang2017effect}, reserve allocation \cite{antenucci2017gas}, and demand response \cite{nikoobakht2020continuous}. In contrast to the steady-state models, these approaches encapsulate the non-linear fluid dynamic equation. The major issue with these works however is that they still require an immense amount of real-time data sharing between two operators, which is not allowed by regulations \cite{zlotnik2016coordinated}. Even if both electricity and natural gas networks are operated by the same entity, the infrastructure to measure and communicate all of the system information is nonexistent. In addition, utilizing these models for a distributed optimization approach such as ADMM does \ul{not} guarantee convergence since these are non-convex models. To deal with the issues regarding privacy and regulatory concerns, robust \cite{he2016robust}, synergistic \cite{wen2017synergistic}, decentralized \cite{wu2020decentralized}, and coordinated \cite{manshadi2018coordinated} scheduling approaches have been proposed, based on the ADMM approach. However, all of these studies utilize steady-state OGF model, which is not able to accurately model gas dynamics, rendering them inappropriate for short-term operation planning periods.

\par This work fills the aforementioned gaps by proposing IV-ADMM, a distributed coordination algorithm whose convergence and accuracy are made possible through a convexified dynamic OGF problem \cite{bayani2022natural}. This enables the electricity and natural gas operators to solve the coordination problem independently, with minimal communication and no regulatory concerns. The only shared information required by this method is the schedule for the generation (or equivalently, the fuel demand) of the gas-fired units. This approach does not conflict with the privacy of each entity, as their objective function remains known only to themselves.

\par
\begin{figure}[h!]
\vspace{-1.2cm}
    \centering
    \includegraphics[width=0.95\linewidth]{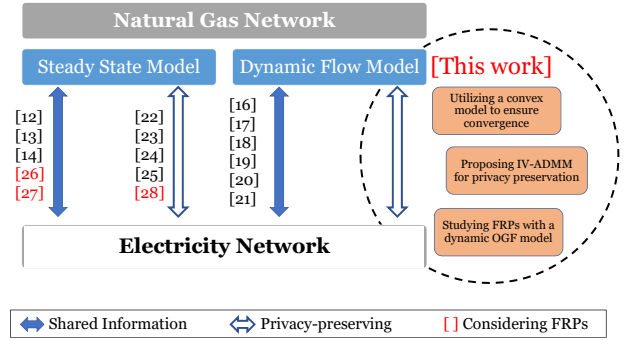}
    \vspace{-1.2cm}
    \caption{Contributions of the current article compared to existing literature} 
    \label{fig:contributions}
    \vspace{-0.75cm}
\end{figure}

 Finally, only a limited count of research work considers FRPs in the context of integrated electricity and natural gas networks. Authors in \cite{mirzaei2019integration} propose a robust scheduling model based on information-gap decision theory in the presence of FRPs. Another approach for coordination of electricity and natural gas networks considering FRPs is presented in \cite{zhang2016electricity}. However, both of these works consider a steady-state gas flow model with a shared information approach. A two-stage method for scheduling of integrated electricity and natural gas networks including FRPs is presented in \cite{song2020two} that does not have privacy concerns. However, the OGF model utilized here is only a linear approximation of the steady state. Additionally, the presented formulation is non-convex which raises issues regarding convergence and generalization of the proposed scheme. This work also contributes to the literature in terms of exploring FRP model with a dynamic OGF model. Figure \ref{fig:contributions} depicts the gap in the relevant literature and emphasizes this work's contributions.

\subsection{\textbf{Contributions}}
\vspace{-.2cm}
\begin{itemize}[leftmargin=*]

\item A \emph{dynamic} natural gas flow model is employed in the \emph{distributed} coordinated scheduling problem of interconnected electricity and natural gas networks. In the literature, the coordinated operation of electricity and natural gas networks is either studied with steady-state OGF models or centralized approaches when dynamics are employed. 

\item A tight convex relaxation scheme tailored for the natural gas dynamics is adopted to employ the OGF problem in a decentralized framework. The benefits of convexifying the original dynamic OGF equations are manifold. First, it ensures the convergence of the IV-ADMM algorithm. Second, a more efficient conic solver can be utilized to solve the dynamic OGF problem. 
Finally, convex relaxation allows obtaining a high-quality tight and tractable solution with much less computation burden than the original non-co OGF model. Reducing the solve time within each iteration is particularly important for the distributed algorithms to be practical since these algorithms may require several iterations to converge.

\item IV-ADMM, a decentralized scheme for the coordinated operation of interconnected electricity and natural gas network, is introduced. No work in the literature has applied decentralized optimization methods to a convex relaxed dynamic OGF problem. IV-ADMM proves more practical than the intensively used classic ADMM in the relevant literature. By applying inexact minimization and varying penalty parameter techniques, IV-ADMM can reach consensus dramatically faster and in fewer iterations than the classic ADMM. This is especially important in applications involving short-term planning horizons, where operators face a limited time window for day-ahead or hour-ahead operation planning. In addition, by applying the distributed IV-ADMM technique, the privacy of each infrastructure is preserved while requiring limited information sharing.

\item The effects of FRPs on the coordinated operation of interconnected electricity and natural gas network are intensely investigated. FRPs are a means of hedging against generation uncertainties in a highly renewable-integrated electricity network. No work in the literature has analyzed the impact of the FRP model in the presence of a dynamic OGF model. The results suggest that when dynamics are considered, FRPs are not appropriate for dealing with uncertainties in the renewable generation.

\end{itemize} \color{black}
\vspace{-0.4cm}
\section{Problem Formulation}\label{sec:formulation}
The coordinated operation problem of electricity and natural gas networks contains two optimization problems. The problem formulation for each one is discussed here.
\vspace{-0.45cm}
\subsection{Electricity Network Day-Ahead Operation Problem}
The optimization problem for the daily operation of the electricity network is presented in \eqref{eq:elec_side}. The objective function of this problem is presented in \eqref{eq:elec_obj}, where the generation costs are approximated by the quadratic function $F^C$ in the first term. The second term in the objective function penalizes unserved loads by factor $\kappa_E$. The served demand at each bus of the system is defined in \eqref{eq:elec_demand_served}, where the superscript $D$ stands for \textit{demand}. The dispatch of wind and solar generation units is limited by their forecast values, as defined in \eqref{eq:elec_dispatch_wt} and \eqref{eq:elec_dispatch_pv}, respectively. The nodal balance equation is displayed in \eqref{eq:elec_balance}. The power flow model for each line is given in \eqref{eq:elec_dc}, where $x_l$ denotes the reactance of that line. Next, the thermal limit of power flow in each line at each time is enforced in \eqref{eq:elec_pline_limit}.\vspace{-0.15cm}
\begin{subequations} \label{eq:elec_side}
\begin{alignat}{3}
&\underset{p,r,\theta}{\min} \sum_{t\in\mathcal{T}}\sum_{g\in\mathcal{G}}^{}{F^C_g(p^G_{g,t})}+\kappa_{E}\sum_{t\in\mathcal{T}}\sum_{b\in\mathcal{B}}^{}{(P^D_{b,t} -p^D_{b,t})},\;\label{eq:elec_obj}\\
& \nonumber \textit{subject to:}\\
& 0 \leq p^D_{b,t} \leq P^D_{b,t},&\hspace{-4.5cm} \forall b\in\mathcal{B}, t\in\mathcal{T}\label{eq:elec_demand_served}\\
& 0 \leq p^{W}_{w,t} \leq P^{W}_{w,t},&\hspace{-4.5cm} \forall w\in\mathcal{W}, t\in\mathcal{T}\label{eq:elec_dispatch_wt}\\
& 0 \leq p^{V}_{v,t} \leq P^{V}_{v,t},&\hspace{-4.5cm} \forall v\in\mathcal{V}, t\in\mathcal{T}\label{eq:elec_dispatch_pv}\\
&\begin{aligned}
\sum_{g\in\mathcal{G}_b}^{}{p^G_{g,t}}+\sum_{v\in\mathcal{V}_b}^{}{p^{V}_{v,t}}+\sum_{w\in\mathcal{W}_b}^{}{p^{W}_{w,t}}=p^D_{b,t}\hspace{1.8cm}\\+\sum_{l\in\mathcal{L}_b^{fr}}{p^L_{l,t}}-\sum_{l\in\mathcal{L}_b^{to}}{p^L_{l,t}}, \hspace{1.8cm}\forall b\in\mathcal{B}, t\in\mathcal{T}\end{aligned}\label{eq:elec_balance}\\
& p^L_{l,t}\cdot x_{l} = \theta_{{b}^{fr}_{l},t} - \theta_{{b}^{to}_{l},t}, &\hspace{-4.5cm} \forall l\in \mathcal{L}, t\in\mathcal{T}\label{eq:elec_dc}\\
&-\overline{P}^L_{l} \leq p^L_{l,t}\leq \overline{P}^L_{l}, &\hspace{-4.5cm}  \forall\;l\in \mathcal{L}, t\in\mathcal{T}\label{eq:elec_pline_limit}\\
& \sum_{g\in\mathcal{G}}^{}{r^{up}_{g,t}} \geq R^{up}_{t} ,&\hspace{-4.5cm} \forall g\in \mathcal{G}, t\in\mathcal{T}\label{eq:flex_ramp_up}\\
& \sum_{g\in\mathcal{G}}^{}{r^{dn}_{g,t}} \geq R^{dn}_{t} ,&\hspace{-4.5cm} \forall g\in \mathcal{G}, t\in\mathcal{T}\label{eq:flex_ramp_dn}\\
& \underline{p}^G_g \leq p^G_{g,t} \leq \overline{p}^G_g,&\hspace{-4.5cm} \forall g\in \mathcal{G}, t\in\mathcal{T}\label{eq:elec_p_limit}\\
& \underline{r}^{up}_{g,t} \leq r^{up}_{g,t} \leq \overline{r}^{up}_{g,t},&\hspace{-4.5cm} \forall g\in \mathcal{G}, t\in\mathcal{T}\label{eq:ramp_up_limit}\\
& \underline{r}^{dn}_{g,t} \leq r^{dn}_{g,t} \leq \overline{r}^{dn}_{g,t},&\hspace{-4.5cm} \forall g\in \mathcal{G}, t\in\mathcal{T}\label{eq:ramp_dn_limit}\\
& \underline{p}^G_g \leq p^G_{g,t}+r^{up}_{g,t} \leq \overline{p}^G_g,&\hspace{-4.5cm} \forall g\in \mathcal{G}, t\in\mathcal{T}\label{eq:elec_rup_limit}\\
& \underline{p}^G_g \leq p^G_{g,t}-r^{dn}_{g,t} \leq \overline{p}^G_g,&\hspace{-4.5cm} \forall g\in \mathcal{G}, t\in\mathcal{T}\label{eq:elec_rdn_limit}
\end{alignat}
\vspace{-0.45cm}
\end{subequations}
\par The FRP model as presented in \cite{wang2014flexible} is utilized to deal with uncertainties in renewable generation availability. According to this model, generation units offer flexible ramping up/down products at each time step. The offered FRPs are later realized in the real-time operation to match the net demand. In other words, FRPs can be regarded as reserve ramping services. The summation of both up and down FRP values offered by all units at each time step should exceed the minimum required ramping service denoted by $R$, as stated in \eqref{eq:flex_ramp_up} and \eqref{eq:flex_ramp_dn}, respectively. Equations \eqref{eq:elec_p_limit}, \eqref{eq:ramp_up_limit}, and \eqref{eq:ramp_dn_limit} set the lower and upper bounds for the generation and ramp up/down values of all units, respectively. Also, according to \eqref{eq:elec_rup_limit} and \eqref{eq:elec_rdn_limit}, each unit at each time cannot offer FRP that will cause them to violate their minimum/maximum generation limits.
\vspace{-0.45cm}
\subsection{Natural Gas Network Dynamic OGF Problem}
The daily operation problem of the natural gas network with dynamics is presented in \eqref{eq:gas_side}. We suppose there are two types of loads in the natural gas system, heat demand, and fuel demand. The heat demand must be met at all times, while the fuel demand of gas-fired units has a lower priority and can be shed, depending on the system conditions. \vspace{-0.15cm}
\begin{subequations} \label{eq:gas_side}
\begin{alignat}{3}
&\min\sum_{t\in\mathcal{T}}\sum_{s\in\mathcal{S}}^{}{\xi_{s.t} q_{s,t}^{S}}+\kappa_{S}\sum_{t\in\mathcal{T}}\sum_{u\in\mathcal{U}}{(F^U_u(p^U_{u,t})-d^U_{u,t})},\;\label{eq:gas_side_obj}\\
& \nonumber \textit{subject to:}\\
&0 \leq d^U_{u,t}\leq F^U_u(p^U_{u,t}),\hspace{1cm}\forall u\in\mathcal{U}, t\in\mathcal{T} :\underline{\tau}^U_{u,t},\overline{\tau}^U_{u,t}\label{eq:gas_side_served}\\
&\begin{aligned}\sum_{s\in\mathcal{S}_j}q_{s,t}^{S}+\sum_{p\in\mathcal{P}^{to}_j} f^{t}_{p,{n}_p}-\sum_{p\in\mathcal{P}^{fr}_j} f^{t}_{p,{1}}=\sum_{h\in\mathcal{H}_j}d^H_{l,t}\hspace{1cm}\\+\sum_{u\in\mathcal{U}_j}d^U_{u,t},\hspace{2cm} \forall j \in \mathcal{J} , t\in\mathcal{T} :\sigma^L_{j,t}\label{eq:gas_side_balance}\end{aligned}\\
& \pi^J_{j,t} = \pi_{p,{n}_p}^{t},\hspace{1.7cm}\forall j \in \mathcal{J},p\in \mathcal{P}^{to}_j, t\in\mathcal{T}: \sigma^{to}_{p,t}\label{eq:gas_side_pr_to}\\
& \pi^J_{j,t} = \pi_{p,{1}}^{t},\hspace{1.3cm}\forall j \in \mathcal{J}\setminus\mathcal{C},p\in \mathcal{P}^{fr}_j, t\in\mathcal{T}: \sigma^{fr}_{p,t}\label{eq:gas_side_pr_from}\\
& \pi^J_{j,t} \leq \pi_{p,1}^{t} \leq \Gamma.\pi^J_{j,t}, \hspace{0cm}\forall j \in \mathcal{C},p\in \mathcal{P}^{fr}_j, t\in\mathcal{T}: \underline{\tau}^{C}_{p,t}, \overline{\tau}^{C}_{p,t}\label{eq:gas_side_pr_comp}\\
&\underline{\pi}_j^{J}\leq \pi^J_{j,t} \leq \overline{\pi}_j^{J},\hspace{1.85cm}\forall j \in \mathcal{J}, t\in\mathcal{T}: \underline{\tau}^J_{j,t},\overline{\tau}^J_{j,t} \label{eq:gas_side_prj_limit}\\
&\underline{q}_s^{S} \leq q_{s,t}^{S} \leq \overline{q}_s^{S},\hspace{2.0cm}\forall s \in \mathcal{S} , t\in\mathcal{T}: \underline{\tau}^S_{s,t},\overline{\tau}^S_{s,t}\label{eq:gas_side_gs_limit}\\
&\begin{aligned} \dfrac{\pi^{t+1}_{p,m+1}-\pi^{t}_{p,m+1}}{\Delta t} +\dfrac{\pi^{t+1}_{p,m}-\pi^{t}_{p,m}}{\Delta t} 
+\dfrac {f^{t+1}_{p,m+1}-f^{t+1}_{p,m}}{c^p_1\Delta x}\;\;\;\\=0,
\hspace{2.3cm}\forall p \in \mathcal{P}, m\in \mathcal{M}_p, t\in\mathcal{T}:\eta^t_{p,m}\label{eq:gas_side_dyna1}
\end{aligned}\\
&\begin{aligned} \dfrac{\pi^{t+1}_{p,m+1}-\pi^{t+1}_{p,m}}{\Delta x}
+\dfrac {f^{t+1}_{p,m+1}-f^{t}_{p,m+1}}{c^p_2\Delta t}
+\dfrac {f^{t+1}_{p,m}-f^{t}_{p,m}}{c^p_2\Delta t}\;\;\;\;\\
+\dfrac{{f^{t}_{p,m}}^2}{c^p_3\pi^{t}_{p,m}}=0,\hspace{1.5cm}\forall p \in \mathcal{P}, m\in \mathcal{M}_p, t\in\mathcal{T}\label{eq:gas_side_dyna2}\end{aligned}
\end{alignat}
\vspace{-0.35cm}
\end{subequations}
\par As given in \eqref{eq:gas_side_obj}, the objective function of the OGF is to minimize gas production costs (where unit price is linearly approximated by parameter $\xi$), while penalizing shedding of gas-fired units' fuel demand by parameter $\kappa_S$. Inclusion of this penalty in the objective ensures the system will deliver the required fuel for generation units as much as possible. The served fuel demand for each gas-fired unit is modeled by \eqref{eq:gas_side_served}, which is bounded by the fuel consumption (approximated by function $F^U$) of the associated generation unit. In \eqref{eq:gas_side_balance}, the nodal mass flow balance at each natural gas junction is enforced. 
Here, the incoming and outgoing mass flow of the pipes connected to each junction should also be considered. 

\par The pressures at the two end segments of each pipeline are considered equal to the pressures at their respective coupling junctions, based on \eqref{eq:gas_side_pr_to} and \eqref{eq:gas_side_pr_from}. The pressure at compressor junctions is modeled by \eqref{eq:gas_side_pr_comp}. According to this model, compressors can boost the pressure up to $\Gamma$ (a parameter greater than 1) times. The pressure at each junction and the amount of gas output of natural gas sources are limited respectively by \eqref{eq:gas_side_prj_limit} and \eqref{eq:gas_side_gs_limit}. Finally, natural gas dynamic equations are displayed in \eqref{eq:gas_side_dyna1} and \eqref{eq:gas_side_dyna2}. These two equations are obtained by applying the finite difference method to the PDEs that model fluid behavior inside pipelines. As a result, the continuous differential equations are discretized into spatio-temporally sectionalized equations. Here, the length of spatial and temporal sections is denoted by $\Delta x$ and $\Delta t$, respectively. The parameters $c^p_1$,$c^p_2$,$c^p_3$ are separately calculated for each pipe based on its diameter, friction factor, and the speed of sound.

\vspace{-0.25cm}
\section {Solution Method}\label{sec:method}
\vspace{-0.05cm}
In the previous section, we presented the formulation for the operation of electricity \eqref{eq:elec_side} and natural gas \eqref{eq:gas_side} networks. These two networks are coupled with the fuel consumption of gas-fired units, which is a function of these units' generation dispatch. The joint solving of \eqref{eq:elec_side} and \eqref{eq:gas_side} is not pragmatic in real-world applications since solving the presented OGF problem is extremely time-consuming and sharing the required amount of data is impractical. In this section, first, a relaxation approach is applied to make the non-convex dynamic OGF formulation convex. Next, the IV-ADMM approach is utilized to obtain the shared decisions. Presenting the convex relaxed form of the OGF problem with dynamics both ensures the convergence of the IV-ADMM algorithm and reduces the computation burden of the OGF problem.
\vspace{-0.3cm}
\subsection{Convex Relaxation of the Dynamic OGF Problem} 
\vspace{-0.05cm}
The last term in \eqref{eq:gas_side_dyna2} makes the natural gas operation problem \eqref{eq:gas_side} a non-convex problem. This means that solving this problem requires non-linear solvers which neither guarantee obtaining optimal solutions nor a polynomial solution time. First, we apply the lifting operation  ${{f^{t}_{p,m}}^2}/{\pi^{t}_{p,m}}\overset{lift}{\longrightarrow}\gamma^t_{p,m}$ to obtain the relaxed form of the non-linear constraint \eqref{eq:gas_side_dyna2}, presented by \eqref{eq:lifting}.
\begin{equation} \label{eq:lifting}
\begin{aligned} \dfrac{\pi^{t+1}_{p,m+1}-\pi^{t+1}_{p,m}}{\Delta x}
+\dfrac {f^{t+1}_{p,m+1}-f^{t}_{p,m+1}}{c^p_2\Delta t}
+\dfrac {f^{t+1}_{p,m}-f^{t}_{p,m}}{c^p_2\Delta t}\;\;\;\;\;\\
+\dfrac{\gamma^t_{p,m}}{c^p_3}=0,\hspace{1.5cm}\forall p \in \mathcal{P}, m\in \mathcal{M}_p, t\in\mathcal{T}:\zeta^t_{p,m}
\end{aligned}
\end{equation}\vspace{-0.3cm}
\par Solving the relaxed OGF problem comprised of \eqref{eq:gas_side_obj}-\eqref{eq:gas_side_dyna1} and \eqref{eq:lifting} does not necessarily yield feasible solutions, i.e. $f^{t}_{p,m}$ and $\pi^{t}_{p,m}$ cannot be uniquely procured from $\gamma^{t}_{p,m}$. 
To handle this issue, a bi-level programming form is introduced which ensures tightness and feasibility of the procured solutions in the upper level. The lower-level problem consists of the relaxed OGF problem. In order to reach an equivalent single-level problem, the closed-form of the relaxed OGF problem is obtained through its dual form presented in \eqref{eq:dual}. In front of each constraint in \eqref{eq:gas_side}-\eqref{eq:dual}, the corresponding primal/dual variables are displayed.
\begin{subequations} \label{eq:dual}
\begin{alignat}{3}
&\begin{aligned}\max\sum_{t\in\mathcal{T}}\sum_{j\in\mathcal{J}}\sigma^L_{j,t} \cdot (\sum_{h\in\mathcal{H}_j}d^H_{l,t}+\sum_{u\in\mathcal{U}_j}F^U_u(r^{up}_{u,t}) ) \\+ \sum_{t\in\mathcal{T}}\sum_{s\in\mathcal{S}}(\underline{q}_s^{S}\underline{\tau}^S_{s,t}-\overline{q}_s^{S}\overline{\tau}^S_{s,t})
-\sum_{t\in\mathcal{T}}\sum_{u\in\mathcal{U}}{F^U_u(p^U_{u,t}) \cdot \overline{\tau}^U_{u,t}} \\ + \sum_{t\in\mathcal{T}}\sum_{j\in\mathcal{J}}(\underline{\pi}_j^{J}\underline{\tau}^J_{j,t}-\overline{\pi}_j^{J}\overline{\tau}^J_{j,t}), \end{aligned}\label{eq:dual_obj}\\
& \vspace{-.35cm}\nonumber \textit{subject to:}\\
&\underline{\tau}^U_{u,t}-\overline{\tau}^U_{u,t}-\sigma^L_{j_u,t}\leq -\kappa_{S},\hspace{.85cm}\forall u \in \mathcal{U}, t\in\mathcal{T}:d^U_{u,t}\label{eq:dual_served}\\
&\begin{aligned}\underline{\tau}^J_{j,t}-\overline{\tau}^J_{j,t}+\sigma^{to}_{p^{to}_j,t}+\sigma^{fr}_{p^{fr}_j,t}\leq 0, \hspace{3.1cm}\\\forall j \in \mathcal{J}\setminus\mathcal{C}, t\in\mathcal{T}:\pi^J_{j,t}\end{aligned}\label{eq:dual_prj}\\
&\begin{aligned}\underline{\tau}^J_{j,t}-\overline{\tau}^J_{j,t}+\sigma^{to}_{p^{to}_j,t}+\underline{\tau}^{C}_{p^{fr}_j,t}- \Gamma.\overline{\tau}^{C}_{p^{fr}_j,t}\leq 0,\hspace{1.5cm}\\
\forall j \in \mathcal{C}, t\in\mathcal{T}:\pi^J_{j,t}\end{aligned} \label{eq:dual_prj_comp}\\
&\sigma^L_{j,t}\geq 0,\hspace{3.2cm}\forall j \in \mathcal{J}^{fr}_p, t\in\mathcal{T}:f^i_{p,t}\label{eq:dual_fin}\\
&\sigma^L_{j,t}\leq 0,\hspace{3.25cm}\forall j \in \mathcal{J}^{to}_p, t\in\mathcal{T}:f^o_{p,t}\label{eq:dual_fout}\\
&\underline{\tau}^S_{s,t}-\overline{\tau}^S_{s,t}+\sigma^L_{j,t} =\xi_{s,t},\forall s \in \mathcal{S}, j \in \mathcal{J}_s,\hspace{.2cm} t\in\mathcal{T}:q^{S}_{j,t}\label{eq:dual_vgs}\\
&\begin{aligned} \dfrac{\eta^{t-1}_{p,m-1}-\eta^{t}_{p,m-1}}{\Delta t} +\dfrac{\eta^{t-1}_{p,m}-\eta^{t}_{p,m}}{\Delta t} + \dfrac{\zeta^{t-1}_{p,m-1}-\zeta^{t-1}_{p,m}}{\Delta x}\;\;\;\;\\\leq 0, \hspace{2cm} \forall p \in \mathcal{P}, m\in \mathcal{M}_p, t\in\mathcal{T}:\pi^t_{p,m}\label{eq:dual_pressure}
\end{aligned}\\
&\begin{aligned} 
\dfrac {\eta^{t-1}_{p,m-1}-\eta^{t-1}_{p,m}}{c^p_1\Delta x}
+\dfrac {\zeta^{t-1}_{p,m-1}-\zeta^{t}_{p,m-1}}{c^p_2\Delta t}
+\dfrac {\zeta^{t-1}_{p,m}-\zeta^{t}_{p,m}}{c^p_2\Delta t}\;\;\;\;\;
\\ \leq 0, \hspace{2cm} \forall p \in \mathcal{P}, m\in \mathcal{M}_p, t\in\mathcal{T}:f^t_{p,m}\label{eq:dual_mass}\end{aligned}\\
&\dfrac{\zeta^t_{p,m}}{c^p_3}\leq 0,\hspace{1.9cm}\forall p \in \mathcal{P}, m\in \mathcal{M}_p, t\in\mathcal{T}:\gamma^t_{p,m}\label{eq:dual_gamma}
\end{alignat}
\vspace{-0.3cm}
\end{subequations}
\par Since the primal and dual forms of the OGF problem are convex, the primal-dual objective equality constraint can be presented as given in \eqref{eq:primal_dual_strong}. \vspace{-0.3cm}
\begin{equation} \label{eq:primal_dual_strong}
\begin{aligned}\sum_{t\in\mathcal{T}}\sum_{s\in\mathcal{S}}^{}{\xi_{s,t} q_{s,t}^{S}}+\kappa_{S}\sum_{t\in\mathcal{T}}\sum_{u\in\mathcal{U}}{(F^U_u(p^U_{u,t})-d^U_{u,t})} \\ = \sum_{t\in\mathcal{T}}\sum_{j\in\mathcal{J}}\sigma^L_{j,t}\sum_{h\in\mathcal{H}_j} d^H_{l,t}+ \sum_{t\in\mathcal{T}}\sum_{s\in\mathcal{S}}(\underline{q}_s^{S}\underline{\tau}^S_{s,t}-\overline{q}_s^{S}\overline{\tau}^S_{s,t})
\\-\sum_{t\in\mathcal{T}}\sum_{u\in\mathcal{U}}{F^U_u(p^U_{u,t})\overline{\tau}^U_{u,t}}+ \sum_{t\in\mathcal{T}}\sum_{j\in\mathcal{J}}(\underline{\pi}_j^{J}\underline{\tau}^J_{j,t}-\overline{\pi}_j^{J}\overline{\tau}^J_{j,t})
\end{aligned}
\vspace{-0.2cm}
\end{equation}
\par Finally, the relaxation process is completed by forming the bi-level problem shown in \eqref{eq:bilevel}, where the upper-level problem enforces the feasibility and tightness of the relaxed OGF problem, and the lower-level problem 
ensures the optimality of the procured solution. A Second-Order Cone Programming (SOCP) method is used along with relaxation, which is displayed in \eqref{eq:bilevel_soc}. To further tighten the relaxed OGF problem, \eqref{eq:bilevel_obj} is placed as the objective of the upper-level problem. Doing so further enforces the tightness by minimizing $\gamma$ values in the second-order cone. The lower-level problem consists of the constraints in the relaxed OGF problem and its dual counterpart, as well as the primal-dual objective equality constraint. 

\begin{subequations} \label{eq:bilevel}
\begin{alignat}{3}
&\min\sum_{t\in\mathcal{T}}\sum_{p\in\mathcal{P}}\sum_{m\in\mathcal{M}_p}\gamma^t_{p,m}\label{eq:bilevel_obj},\\
& \nonumber \textit{subject to: }  \eqref{eq:gas_side_served}-\eqref{eq:gas_side_dyna1},\eqref{eq:lifting},\eqref{eq:dual_served}-\eqref{eq:dual_gamma}, \eqref{eq:primal_dual_strong}\\
& \hspace{-.35cm}\begin{Vmatrix} 2f^t_{p,m}\\\gamma^t_{p,m}-\pi^t_{p,m}
\end{Vmatrix} \leq \gamma^t_{p,m}+\pi^t_{p,m}, \forall p \in \mathcal{P}, m\in \mathcal{M}_p, t\in\mathcal{T} \label{eq:bilevel_soc}
\end{alignat}
\end{subequations}
\vspace{-0.75cm} 
\subsection{Privacy-Preserving Coordinated Operation via IV-ADMM}
Due to the limitations regarding privacy, regulations, and communication infrastructure, the coordinated operation of coupled electricity and natural gas networks is almost impossible in practice. To address this, we present IV-ADMM, a practical privacy-preserving approach that only requires sharing the dispatch of gas-fired units between the two networks. If the objectives of problems \eqref{eq:elec_side} and \eqref{eq:bilevel} are denoted respectively by $f(x)$ and $g(z)$, the augmented Lagrangian function for solving the coordinated operation problem can be defined as \eqref{eq:lagrangian}. Here, $x$ and $z$ denote the decisions for dispatch of gas-fired units at each step, respectively obtained by the electricity and the natural gas networks. The penalty parameter and the Lagrangian multiplier are also denoted by $\rho$ and $\lambda$, respectively.
\begin{equation}\label{eq:lagrangian}
    L_{\rho}(x,z,\lambda) = f(x)+g(z)+\left \langle \lambda,x-z \right \rangle + \rho/2\begin{Vmatrix} x-z\end{Vmatrix}^2_2
\end{equation}
\par IV-ADMM accelerates the solution process of the distributed optimization problems \eqref{eq:elec_side} and \eqref{eq:bilevel} based on the concepts referred to as \textit{inexact minimization} and \textit{varying penalty parameter}. 

\subsubsection{Inexact Minimization} 
According to the results provided by \cite{eckstein1992douglas}, the algorithm's minimization problems can be solved approximately in the initial iterations and more accurately later on. To this end, we use an exponentially decaying convergence tolerance for the solver, as shown in \eqref{eq:conv_tol}. Here, the tolerance ($\tau$) denotes the relative difference between the objective values of primal and dual. The iteration number in each step of the algorithm is denoted by $k$, and $\alpha$ and $\beta$ are constant parameters.
\begin{equation}
    \label{eq:conv_tol}
    \tau^{k} = \alpha \cdot \beta ^ {-k} 
\end{equation}

\subsubsection{Varying Penalty Parameter}
As a practical measure, the penalty parameter ($\rho$) can be calibrated after each iteration. Consider $r$ and $s$ to respectively denote the residual values for primal and dual at each iteration of the algorithm, defined by \eqref{eq:residual_primal} and \eqref{eq:residual_dual}.
\begin{equation}
    \label{eq:residual_primal}
    r^{k} =  x^{k} - z^{k} 
\end{equation}
\vspace{-.5cm}
\begin{equation}
    \label{eq:residual_dual}
    s^{k+1} = \rho \cdot ( z^{k+1} - z^{k} )
\end{equation}
\par One criteria for IV-ADMM convergence is the value of these residuals, i.e. they should approach zero. By applying the varying penalty parameter scheme displayed in \eqref{eq:varying_penalty}, the convergence is enhanced in practice \cite{boyd2011distributed}. This also makes sure that the initial choice of $\rho$ does not affect IV-ADMM performance.
\begin{equation} \label{eq:varying_penalty}
\rho^{k+1}=\left\{\begin{array}{ll}
\rho^{k} \cdot \delta & \text { if }\left\|r^{k}\right\|_{2} > \omega \left\|s^{k}\right\|_{2} \\
\rho^{k} / \delta & \text { if }\left\|s^{k}\right\|_{2} > \omega \left\|r^{k}\right\|_{2} \\
\rho^{k} & \text { otherwise, }
\end{array}\right.
\end{equation}
\par Here, $\delta$ and $\omega$ are constant parameters greater than one. The idea behind \eqref{eq:varying_penalty} is when the primal residual is very large (i.e. the decisions of the two networks are far apart), $\rho$ is increased, which causes the last term in the Lagrangian function \eqref{eq:lagrangian} be penalized even more. Conversely, if the primal residual becomes very small (i.e. the decisions of the two networks are very close), the penalty parameter is reduced, which eventually reduces the dual residual value by inhibiting fluctuations in the decisions of consecutive iterations.

\par The consensus-based privacy-preserving coordinated operation IV-ADMM algorithm applied to solve the coordination problem of natural gas and electricity networks is displayed in Algorithm \ref{alg_1}. At each step of this algorithm, each network obtains and passes the dispatch of gas-fired units based on the operational constraints and the decision communicated by the other network. This message-passing scheme continues until the decisions of both networks become relatively equal with regard to a certain threshold. After convergence criteria are met, consensus is reached and the solution process is terminated. The input parameters of this algorithm are the maximum number of iterations ($k_{max}$), the initial penalty parameter ($\rho^0$), the convergence threshold ($\epsilon$), and the parameters $\alpha$, $\beta$, $\delta$, and $\omega$.
\vspace{-0.15cm}
\begin{algorithm}[h!] \label{alg_1}
    \SetKwInOut{Input}{Input}
    \SetKwInOut{Output}{Output}
    \SetKwInOut{Initiate}{Initiate}
    \caption{IV-ADMM Algorithm}
    \Input{$k_{max},\;\rho^0,\; \epsilon,\; \alpha,\; \beta,\; \delta, \; \omega$}
    \Output{$\boldsymbol p^U$}
    \Initiate{$\boldsymbol z^0,\boldsymbol\lambda^0 \leftarrow \textbf{0}_{N_U\times N_T}$}
    \For {$k = 0:k_{max}$}
        {
        $\tau^{k} \leftarrow \eqref{eq:conv_tol}$\\
        set solver's tolerance to $\tau^{k}$\\
        $\boldsymbol x^{k+1}\leftarrow \arg \min_x L_{\rho}(\boldsymbol x,\boldsymbol z^k,\boldsymbol\lambda^k),\; \newline \textit{subject to: } \eqref{eq:elec_demand_served}-\eqref{eq:elec_rdn_limit}$ \vspace{.1cm}\\
        $\boldsymbol z^{k+1}\leftarrow \arg \min_z  L_{\rho}(\boldsymbol x^{k+1},\boldsymbol z,\boldsymbol\lambda^k),\newline \textit{subject to: } \eqref{eq:gas_side_served}-\eqref{eq:gas_side_dyna1},\eqref{eq:lifting},\eqref{eq:dual_served}-\eqref{eq:dual_gamma}, \eqref{eq:primal_dual_strong},\eqref{eq:bilevel_soc}$ \vspace{.1cm}\\
        $\boldsymbol \lambda^{k+1}\leftarrow \boldsymbol \lambda^{k}+\rho^{k}(\boldsymbol x^{k+1}-\boldsymbol z^{k+1})$ \\
        $r^k \leftarrow \eqref{eq:residual_primal}$\\
        $s^k \leftarrow \eqref{eq:residual_dual}$\\
        $\rho^{k+1} \leftarrow \eqref{eq:varying_penalty}$\\
        $\boldsymbol h \leftarrow \textbf{0}_{N_U\times N_T}$\\
        \For {$u\in\mathcal{U}$}
            {
            \For{$t\in\mathcal{T}$}
                {                \If{$\dfrac{|x^{k+1}_{u,t}-z^{k+1}_{u,t}|}{(x^{k+1}_{u,t}+z^{k+1}_{u,t})/2}\leq \epsilon$}
                    {$h_{u,t}\leftarrow1$}
                }
             }
            \If{$\boldsymbol h = \textbf{1}_{N_U\times N_T}$}{\textbf{break}
            }
        }
    \end{algorithm}
\vspace{-0.35cm}
\par The IV-ADMM algorithm begins by initializing the natural gas network decision ($\boldsymbol z$) and the Lagrangian multiplier ($\boldsymbol \lambda$) to zero. Before anything, the tolerance $\tau$ of the solver is acquired based on \eqref{eq:conv_tol} and is stored at line 3. At line 4, the electricity network's decision is obtained through minimizing the Lagrangian function, which is passed to the natural gas network. The natural gas network's decisions are then acquired by minimizing the Lagrangian function displayed in line 5. These two steps are basically the same as solving \eqref{eq:elec_side} and \eqref{eq:bilevel}, but with modified objectives so that the decision of the other network is taken into consideration.
At line 6, the Lagrangian multiplier term is acquired based on the difference between the decisions of the two networks. At this stage, the residual values for primal and dual are obtained through \eqref{eq:residual_primal} and \eqref{eq:residual_dual}, respectively, and the penalty parameter for the next iteration is updated based on \eqref{eq:varying_penalty}. At the end of each iteration, the \textit{relative error} of the decisions obtained by both networks is calculated at line 13. Convergence is reached when the relative error for each unit's decision at all times becomes less than $\epsilon$. At this stage, all of the decisions of both networks are equal, and the two systems have agreed on a consensus.\vspace{-0.35cm}

\section {Results and discussion}\label{sec:result}
The results of applying the proposed methodology to coordinate the operation of an interconnected electricity and natural gas network are investigated in several cases. In addition to the sample coupled networks, the coordinated operation of a large coupled natural gas and electricity network is shown to demonstrate the scalability of the proposed method. In all of our experiments, the simulation period is 24 hours, segregated into 5-minute time intervals. The simulations are performed on a PC with a Core i7 3.6 GHz CPU using the Julia \cite{bezanson2017julia} language and the JuMP \cite{DunningHuchetteLubin2017} environment. Optimizations are solved using Gurobi \cite{gurobi}, and the solver parameter ``BarQCPConvTol" is selected for setting the tolerance $\tau$.
\begin{figure}[b!]
    \vspace{-0.75cm}
    \centering
    \includegraphics[width=0.8\linewidth]{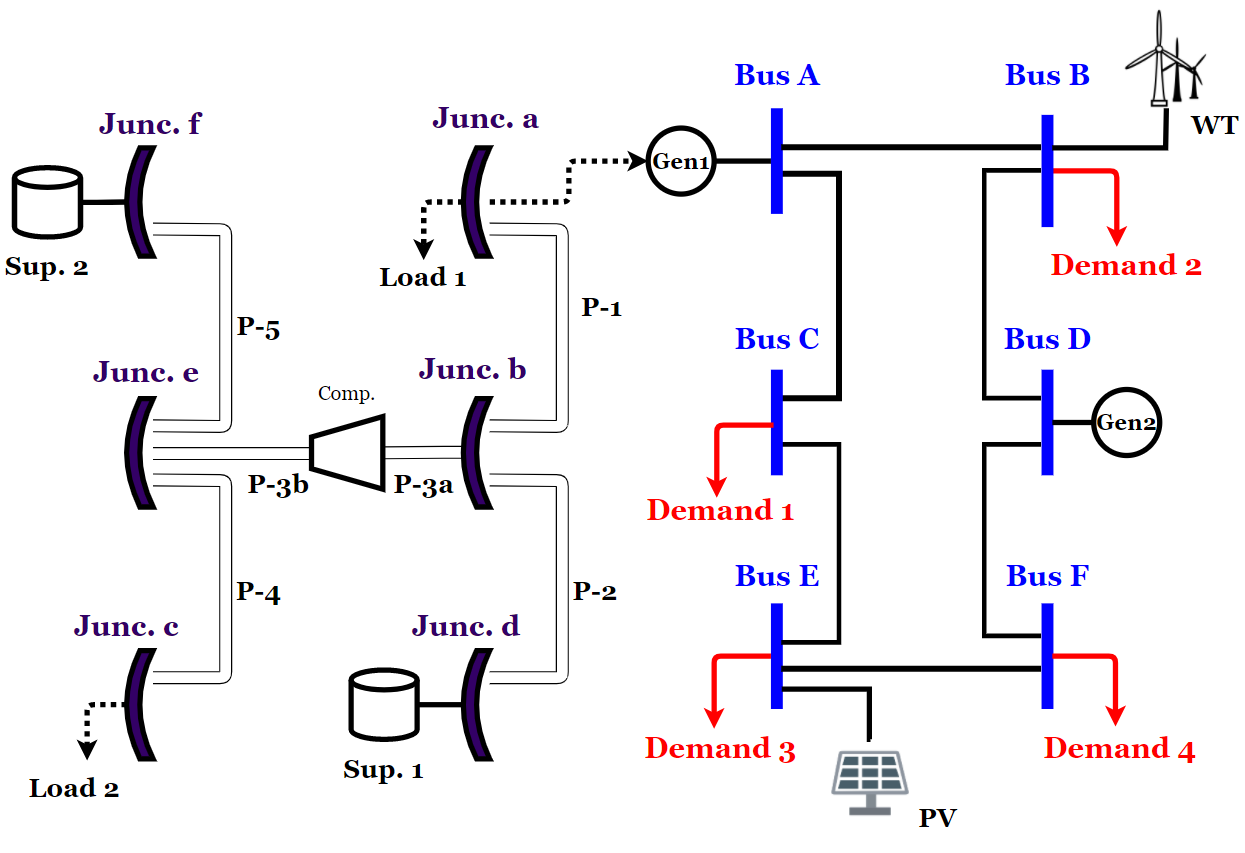}
    \vspace{-0.45cm}
    \caption{6-Junction natural gas network coupled with 6-bus electricity network}
    \label{fig:coupled_6}
    \vspace{-0.35cm}
\end{figure}
\vspace{-0.55cm}\subsection{\textbf{Case 1: Day-Ahead Operation}} \vspace{-0.15cm} \label{results:case1}
Here, the day-ahead coordinated operation problem of a highly RES integrated 6-bus electricity network interconnected with a 6-junction natural gas network is considered. 
The peak daily electricity load of the system is 360 MW. As displayed in Fig. \ref{fig:coupled_6}, two generation units provide the required demand for the network. One generation unit is gas-fired (Gen. 1) and the other one (Gen. 2) is thermal. The specifications of these units are displayed in Table \ref{table:small_network_data}. 
In this case, due to the high penetration level of renewable generations, an hourly FRP requirement equivalent to $40\%$ of the peak load is considered for all times. This means at each 5-minute interval, the combined up and down FRP of generation units should exceed 12 MW at all times.
\setlength\tabcolsep{2.9pt} 
\begin{table}[h!] \vspace{-0.25cm} \centering
\caption{generation unit data in 6-bus electricity network} \vspace{-0.25cm}
\resizebox{\linewidth}{!}{
\begin{tabular}{ccccc} \hline\hline
Unit & $P_{min}$ (MW) & $P_{max}$ (MW) & Ramp rate (MW/hr) &  5-min ramp rate  \\ \hline 
1 & 0 & 150 & 125 &  10.42 \\
2 & 50 & 225 & 75 &  6.25 \\\hline \hline
\label{table:small_network_data}
\end{tabular}
} 
\vspace{-0.55cm}
\end{table}

\begin{figure}[h!] \centering
    \vspace{-0.15cm}
    \includegraphics[width=0.9\linewidth]{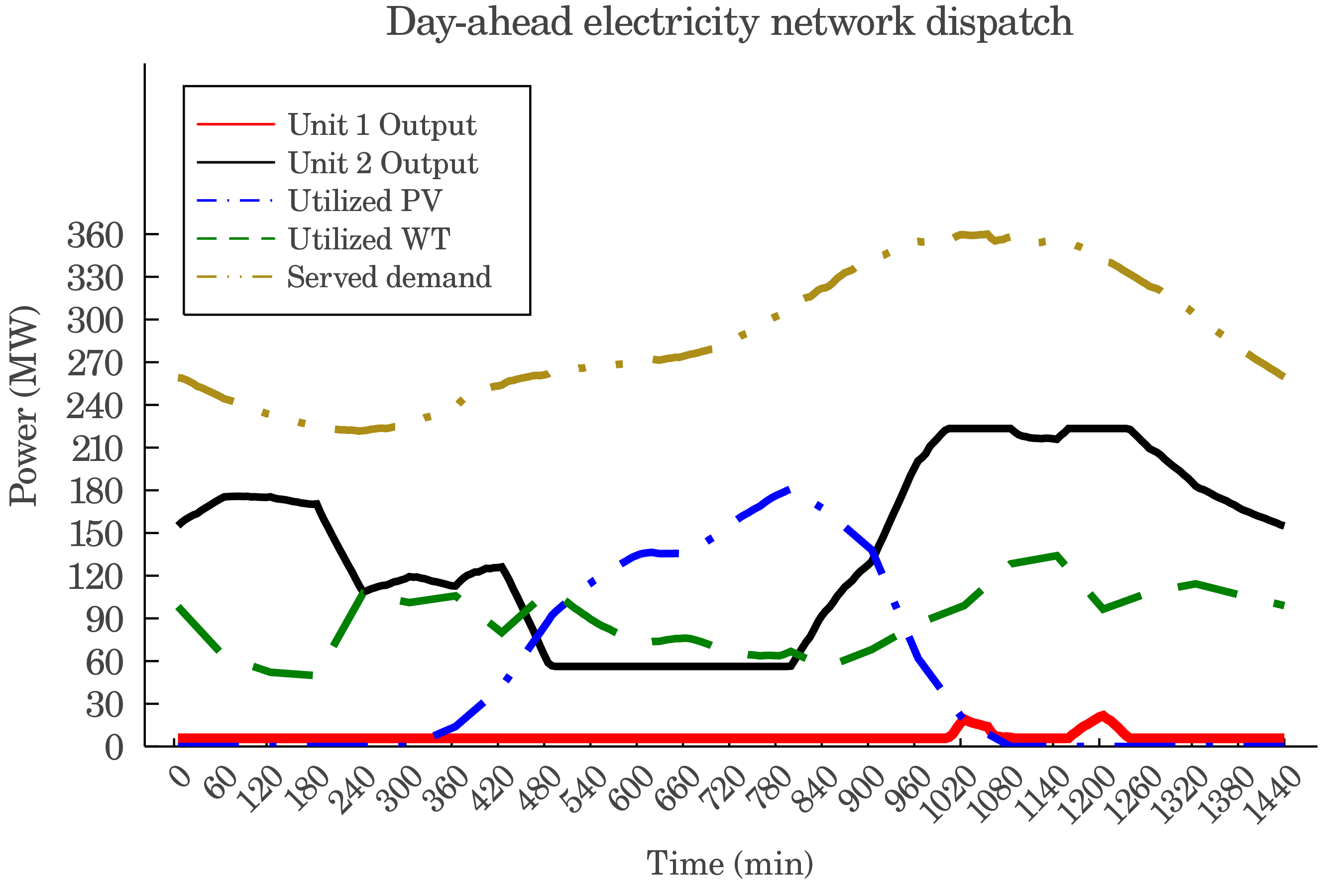}
    \vspace{-0.35cm}
    \caption{Electric power dispatch of units in the 6-bus power network}
    \label{fig:case1_dispatch}
    \vspace{-0.55cm}
\end{figure}

\subsubsection{Coordinated Operation Schedule}
\par According to the day-ahead dispatch results presented in Fig. \ref{fig:case1_dispatch}, the interconnected system is prepared to serve all of the predicted electricity demand. In this case, the gas-fired generator (unit 1) will provide a small non-zero generation throughout the day to compensate the deficiency in FRP requirement provided by the other unit. Here, the generated power between RESs and the non-gas-fired generator (unit 2) is almost sufficient for meeting the required daily demand. The only exception happens during the peak load period, which the participation of the gas-fired is also required.

\par
\begin{figure}[h!]
    \vspace{-0.35cm}
    \centering
    \includegraphics[width=0.9\linewidth]{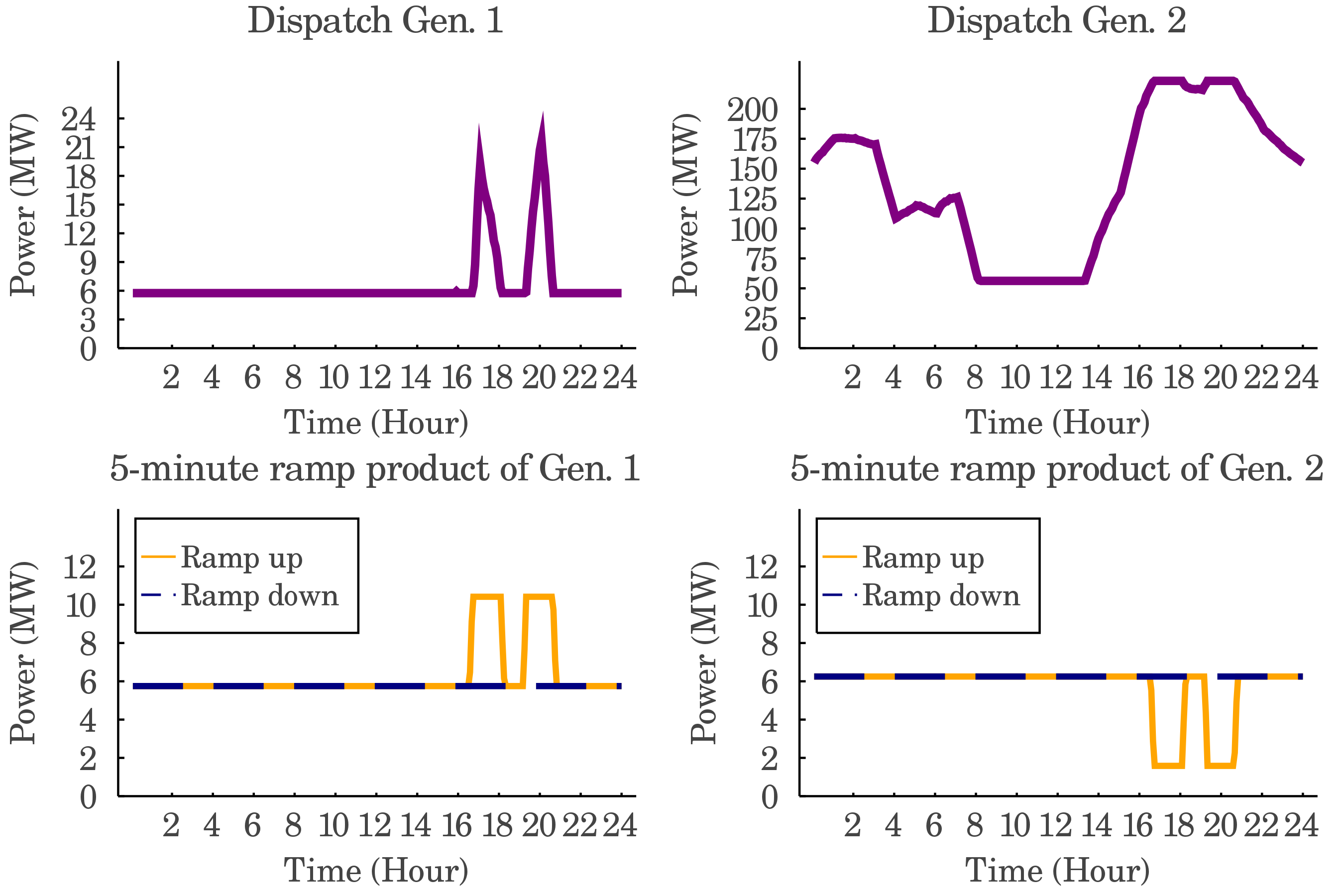}
    \vspace{-0.35cm}
     \caption{Dispatch (top) and FRP (bottom) results of units in Case 1}
   \label{fig:case1_ramp}
    \vspace{-0.25cm}
\end{figure}

 It is noticed that the dispatch of the gas-fired unit is non-zero even during off-peak hours. This is because this unit has to assist the other unit in providing the FRP requirement. The breakdown results for the dispatch and ramp products of generation units in Case 1 are displayed in Fig. \ref{fig:case1_ramp}. Here, unit 2 cannot offer an FRP of more than 6.25 MW per 5-minute interval. To meet the 12 MW FRP requirement of the system, unit 1 has to provide at least 5.75 MW FRP at all times. According to \eqref{eq:elec_rdn_limit}, the output power of each unit should be greater than its flexible \textit{ramping-down} product. This is why unit 1 is consistently generating no less than 5.75 MW throughout the day. During the period 4:45'-8:35' in the evening, the output of solar units drops, and at the same time, a surge occurs in the net demand. It is only during this period that unit 1's output contributes solely to meeting the electricity demand since unit 2 has reached its maximum capacity.
\subsubsection{Comparing the Performance of IV-ADMM with the Classic ADMM}
\par
\begin{figure}[h!]
    \vspace{-0.35cm}
    \centering
    \includegraphics[width=1\linewidth]{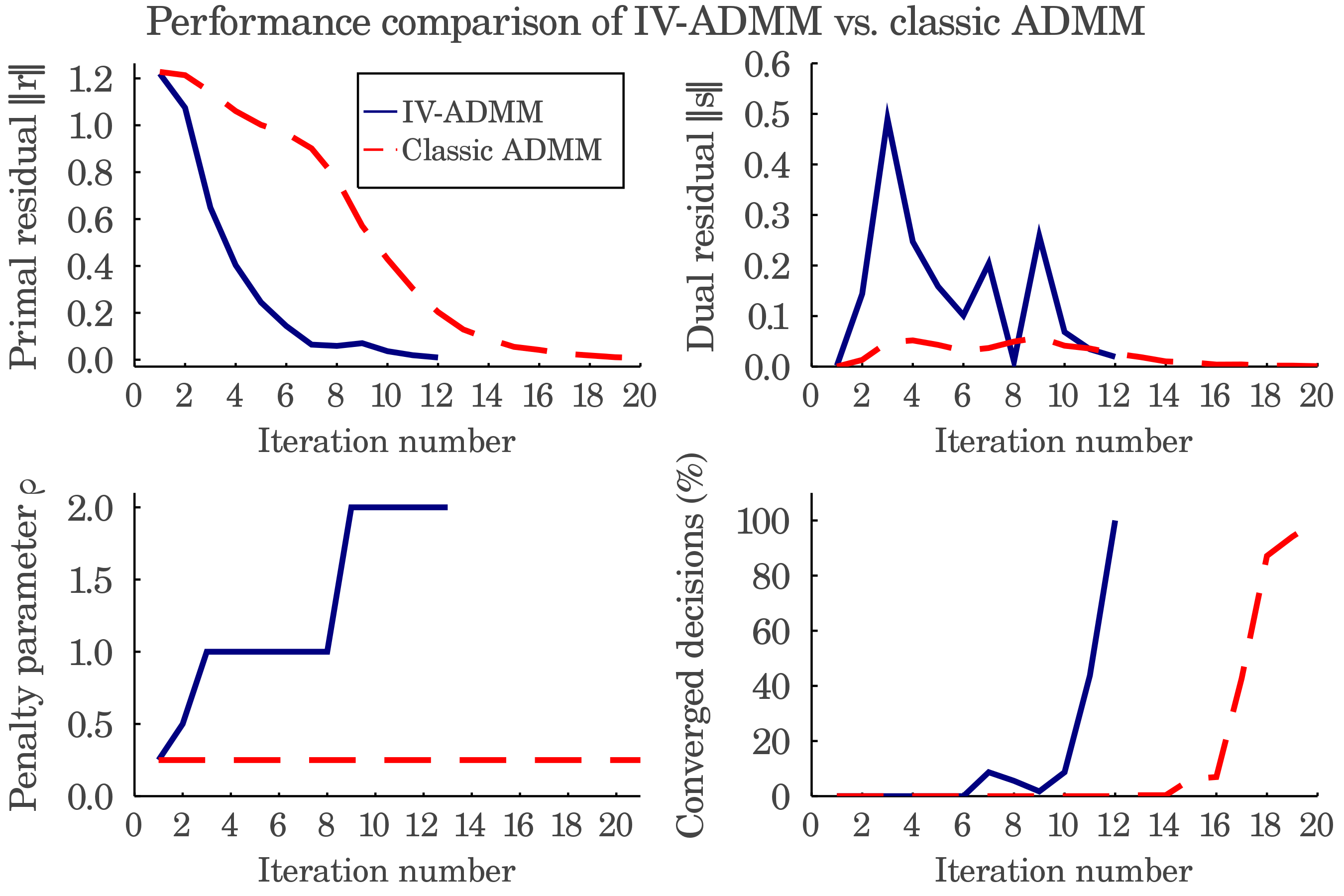}
    \vspace{-0.75cm}
    \caption{Performance comparison of IV-ADMM vs. classic ADMM in Case 1}
    \label{fig:case1_admm_comparison}
    \vspace{-0.25cm}
\end{figure}

 
The results show IV-ADMM is able to reach consensus considerably faster and with a lower number of iterations compared with the classic ADMM. It is observed in Fig. \ref{fig:case1_admm_comparison} that IV-ADMM converges in 12 iterations while classic ADMM converges in 20 iterations. From the run time standpoint, classic ADMM takes 623.5 seconds to converge, while IV-ADMM ends in 248.2 seconds. IV-ADMM reaches consensus in all decisions by requiring only 60\% and 40\% of classic ADMM's iteration count and run time, respectively. The reason that these numbers are disproportionate is that in the initial iterations, IV-ADMM stops the solver sooner than classic ADMM, leading to a considerably shorter iteration duration. The other contributing factor to IV-ADMM's faster run time is the varying $\rho$ concept. Based on the primal and dual residual values obtained by the IV-ADMM algorithm, it is observed that the first condition in \eqref{eq:varying_penalty} is satisfied three times (at iterations 1, 2, and 8). Consequently, the value of $\left\|r^{k}\right\|$ descends rapidly, which accelerates the convergence of decisions. The values of the parameters used to run Alg. \ref{alg_1} are as follows: $\epsilon = 0.02$, $\rho^0=0.25$, $\alpha=0.1$,\; $\beta=2.4$, $\delta=2$, and $\omega=3.8$.

\subsubsection{The interaction of FRPs and the natural gas dynamics}
\par
\begin{figure}[h!]
    \vspace{-0.35cm}
    \centering
    \includegraphics[width=0.9\linewidth]{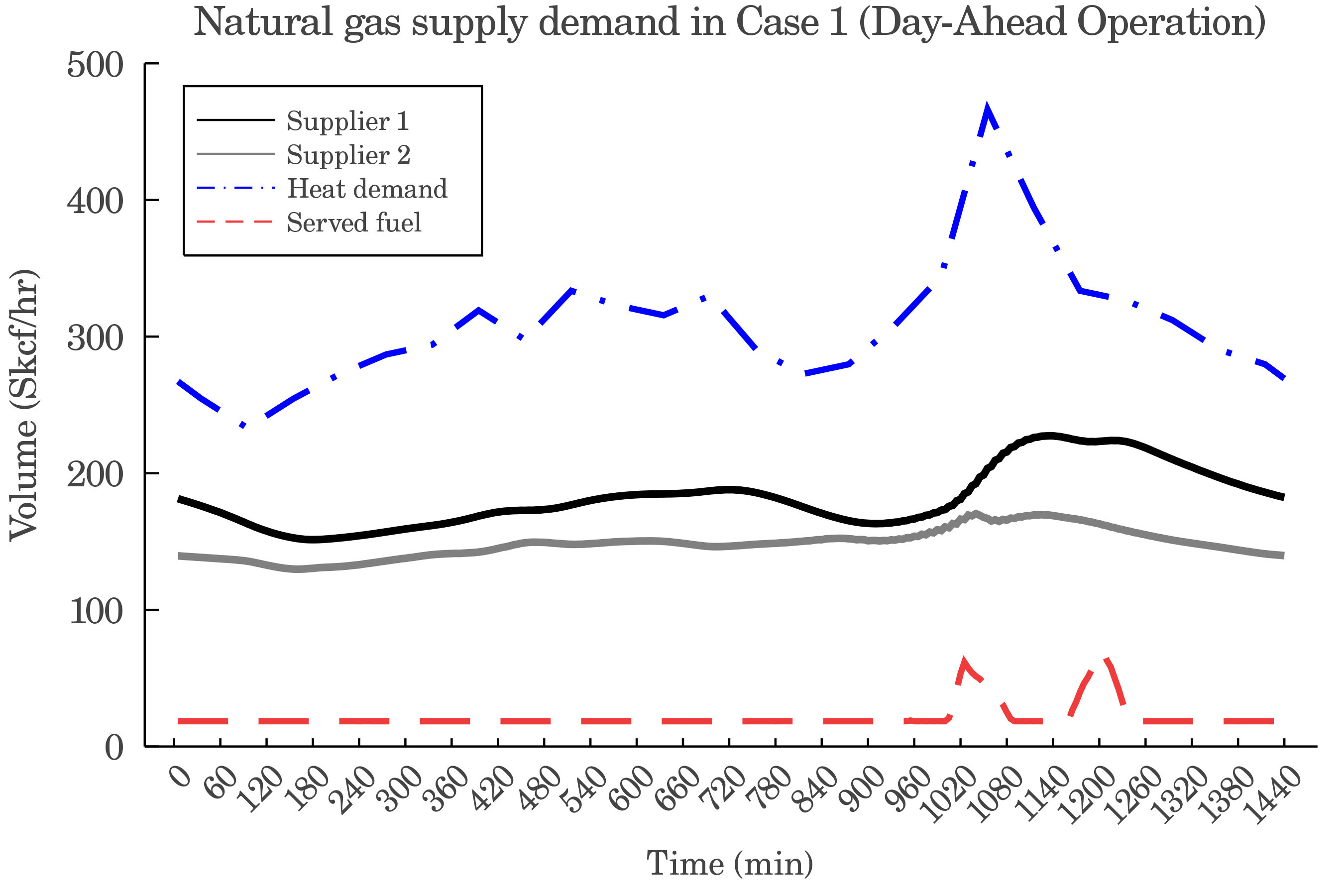}
    \vspace{-0.35cm}
    \caption{Natural gas network supply demand curve in Case 1}
    \label{fig:case1_gas_dispatch}
    \vspace{-0.35cm}
\end{figure}

 Variations in the demand within the natural gas network are met with adjustments in the pressure of junctions rather than changes in the output of suppliers. This is because the transportation of natural gas fluid inside pipelines is not instantaneous, and both spatial and temporal dimensions are crucial to the dynamic OGF model. 
Fig. \ref{fig:case1_gas_dispatch} shows the daily supply demand curve of the natural gas network in Case 1. Although an arbitrarily highly varying heat demand is considered for the system, it is noticed that the changes in the supplier outputs cannot account for the variations in the served demand.
\par
\begin{figure}[h!]
    \vspace{-0.35cm}
    \centering
    \includegraphics[width=0.9\linewidth]{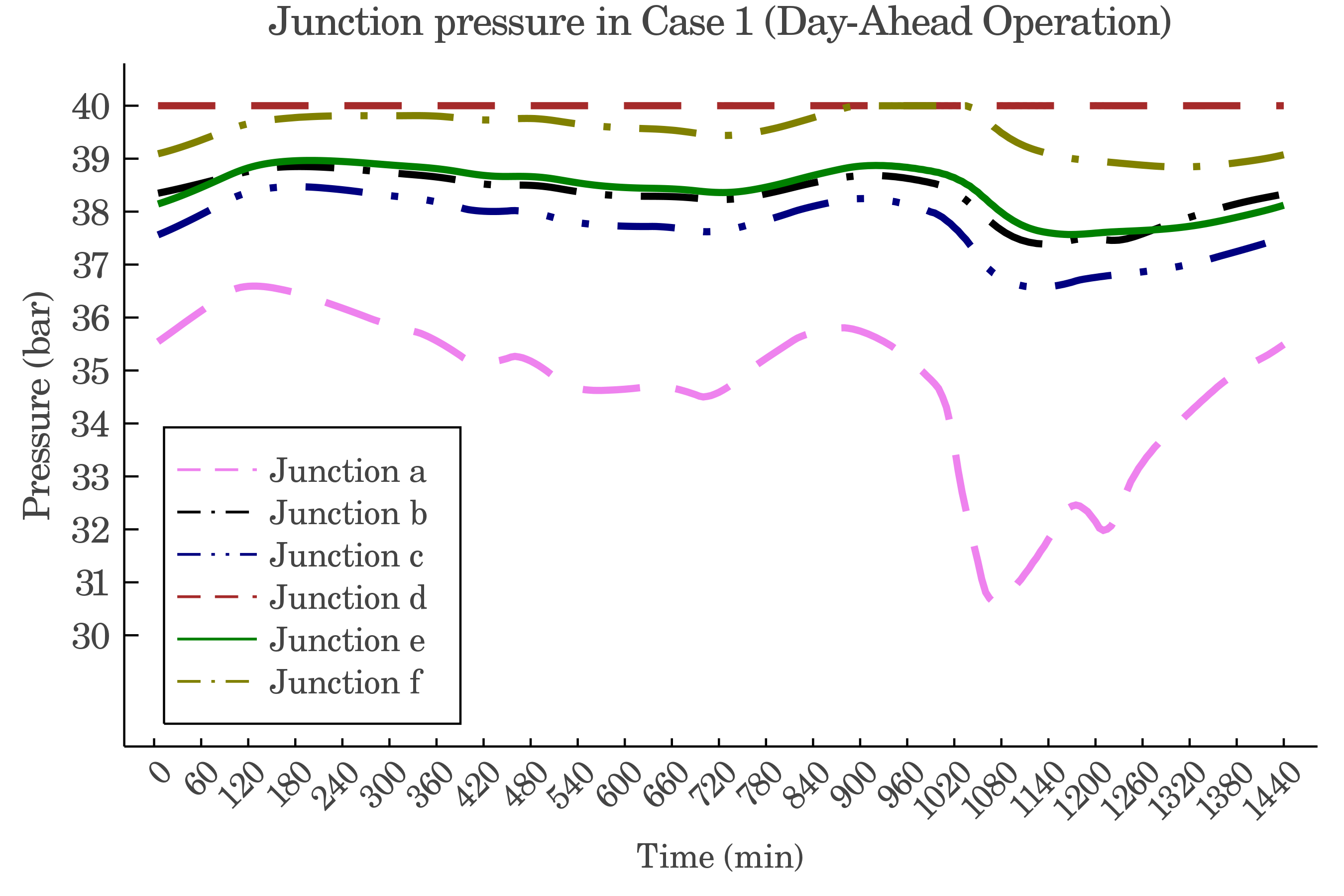}
    \vspace{-0.35cm}
    \caption{Pressure of natural gas network junctions in Case 1}
    \label{fig:case1_pressure}
    \vspace{-0.15cm}
\end{figure}

 The results for the pressure of network junctions throughout the day displayed in Fig. \ref{fig:case1_pressure} indicate that it is the pressure that is accountable for matching the demand alterations. By comparing the pressure curves of junctions `d' and `a', where supplier `1' and load `1' of the natural gas network are placed respectively, it is noticed that the pressure at the supplier junction remains steady at all times. However, the pressure at the demand junction is adjusting in response to the changes in the demand curve (in the opposite direction). This suggests that the \textit{difference} between the pressures at corresponding junctions is the most prominent factor affecting gas dynamics, and it should be investigated closely to grasp a better understanding of the governing dynamics of natural gas.
\vspace{-0.15cm}
\subsubsection{Verification of the Proposed Convex Relaxation Scheme}
Consider the SDP form equivalent of the SOC constraint \eqref{eq:bilevel_soc}, displayed in \eqref{eq:sdp_matrix}. A tight convex relaxation is achieved if this matrix is rank-1 in all spatio-temporal segments, . In this case, the lifted variable $\gamma$ can be uniquely retrieved from the lifting operation introduced in part A of section III.
\begin{equation}\label{eq:sdp_matrix}
\begin{aligned}
    \begin{bmatrix}
    pr^{t}_{p,s} & m^{t}_{p,s}\\ 
    m^{t}_{p,s} & \gamma^t_{p,s}
    \end{bmatrix} , \;\;\;\;\forall p \in \mathcal{P}, s \in \mathcal{S}_p, t\in\mathcal{T}
\end{aligned}
\end{equation}
\par We use tightness ratio \cite{soofi2020socp} to measure the tightness of the convex relaxation. For each matrix, this value is defined as the logarithmic ratio of its largest and second-largest eigenvalues. In the last iteration of Case 1, the average tightness ratio for all spatio-temporal matrices was equal to 6.3. It means that the lifted variable can be recovered with an error of less than $10^{-6}$. This suggests that the obtained answer by the applied relaxation scheme is tight and feasible.

\vspace{-0.25cm}
\subsection{\textbf{Case 2: FRPs Realization in the Real-Time Operation}}
\par In Case 2, we put the day-ahead results into test by considering a scenario where every predicted day-ahead value is realized exactly as predicted in real-time, except for the output of WT units. According to the results of the day-ahead operation problem presented in Case 1, if in the real-time operation scenario, the realized net demand deviates from the predicted value by up to $40\%$ of the peak load, the total electricity demand \textit{should} be fully served. Here, the considered curve for the WT output is such that it does not deviate more than the FRP requirement designed in Case 1.
\begin{figure}[h!]
    \vspace{-0.35cm}
    \centering
     \includegraphics[width=0.9\linewidth]{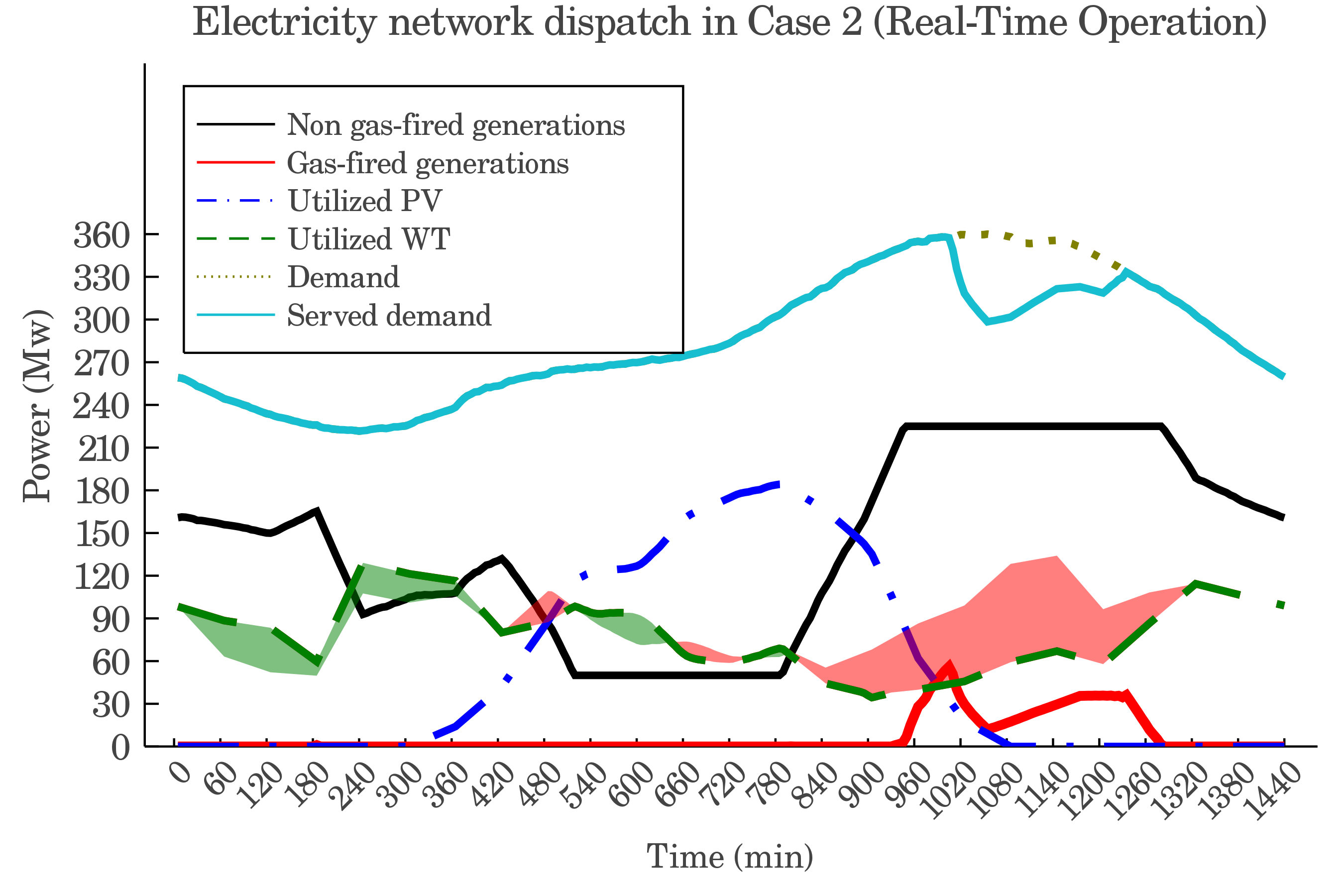}
    \vspace{-0.35cm}
    \caption{Electric power dispatch of units in the 6-bus power network}
    \label{fig:case2_dispatch}
   \vspace{-0.15cm}
\end{figure}
\par Although the day-ahead operation included the FRP model to hedge against uncertainties, the results of the real-time scenario indicate that due to the incapability of the natural gas network to deliver the required fuel for the gas-fired unit's generation, $2.03 \%$ of the daily electricity demand is shed. The daily dispatch of the electricity network in Case 2 is displayed in Fig. \ref{fig:case2_dispatch}. Here, the shaded green and red areas respectively denote the increase and decrease in the amount of the WT unit's real-time dispatch compared with the predicted amounts in the day-ahead scenario (Case 1). It is observed that during off-peak hours, the change in the output of the WT unit is reflected in the output of the non-gas-fired unit. However, during the evening, when the output of the non-gas-fired unit has reached its maximum capacity and consumer demand is still surging towards its peak, the drop in WT generation should be met with the rise in gas-fired units' output. Interestingly, it is noticed that the natural gas network dynamics do not allow for the complete fuel delivery to the gas-fired unit. This case shows why FRPs cannot be trusted as a reliable resort for matching real-time fluctuations, as their impact on the natural gas network is not reflected in the formulations. The advantage of utilizing the exact dynamic gas model in this work could be illustrated in this case, as steady-state models are not capable of correctly predicting natural gas behavior on such occasions.

\subsection{\textbf{Case 3: Remedy Natural Gas Network Limitations}}
\par Case 3 is designed to better understand the governing dynamics of natural gas. Here, every operational condition is the same as that of Case 2. But this time, we suppose that the pipeline diameters are increased by 20\%. Expanding the volume of natural gas pipelines increases the pipeline gas storage capacity at the same pressure. In \eqref{eq:gas_side_dyna1} and \eqref{eq:gas_side_dyna2}, by increasing the parameters $c_1^p,\;c_2^p,\; \text{and }c_3^p$, more mass is able to flow through pipelines. Since expansion of the  pipe diameter increases all of these parameters, it is expected that the delivery issues confronted in Case 2 no longer exist in Case 3, and all of the fuel required by the gas-fired unit will be delivered.
\par
\begin{figure}[h!]
    \vspace{-0.35cm}
    \centering
    \includegraphics[width=1\linewidth]{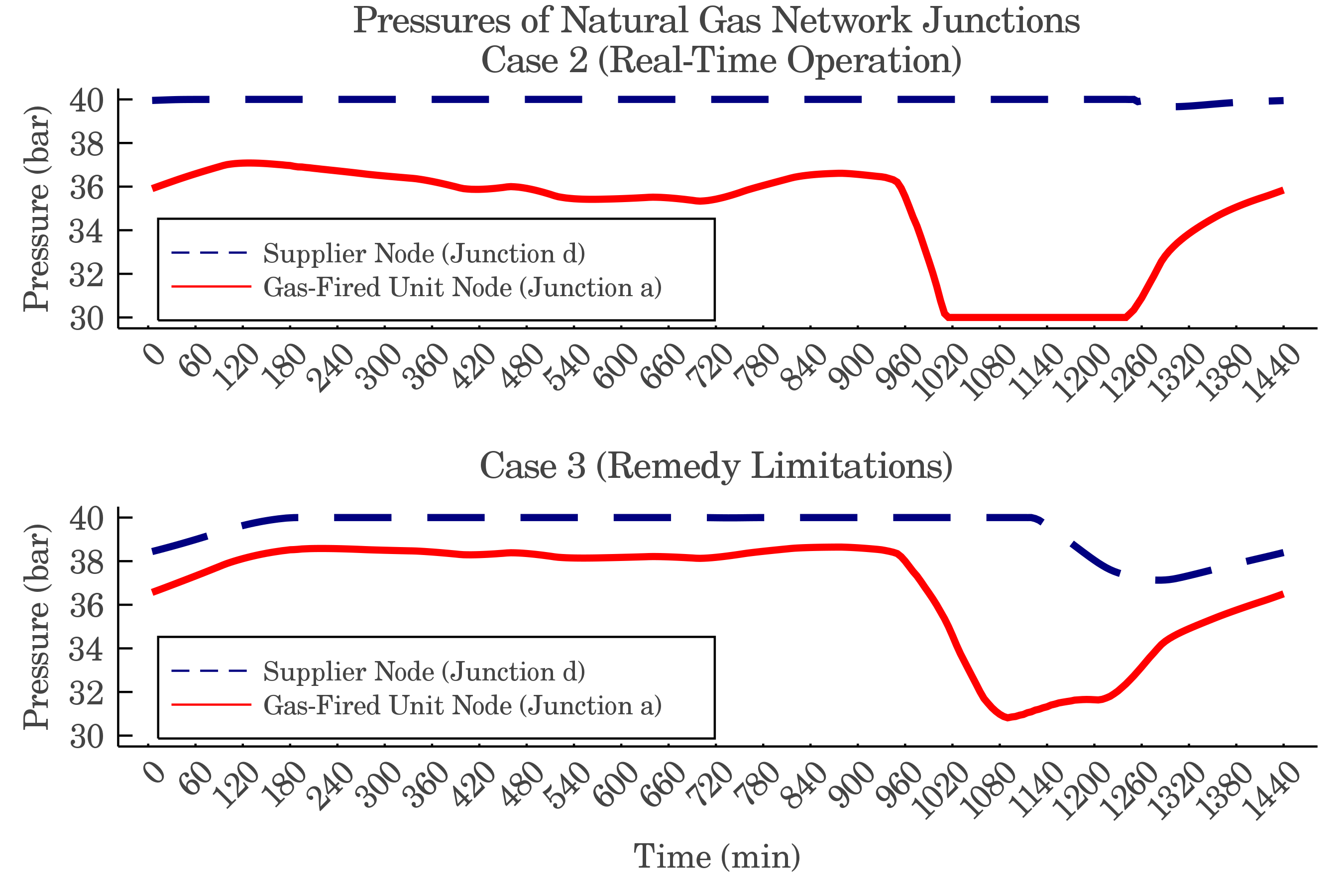}
    \vspace{-0.85cm}
    \caption{Pressure at junctions of gas network in Cases 2 (top) and 3 (bottom)}
    \label{fig:case3_pressure}
    \vspace{-0.35cm}
\end{figure}

 Increasing the pipe diameter in Case 3 leads to more gas delivery at the same pressure gap between the two junctions. Consequently, all of the fuel demand is supplied and the electricity demand is totally met in Case 3. The pressure of natural gas network junctions at Cases 2 and 3 are illustrated in Fig. \ref{fig:case3_pressure}. According to \eqref{eq:gas_side_dyna2}, by increasing the pressure difference between two points in the system, more mass can be stored inside pipelines. This means one way to deal with the increased natural gas demand at a junction is to reduce its pressure. However, operational constraints do not allow pressure to drop below a certain threshold. In the introduced test system, the pressure's lower and upper bounds are set at $30$ and $40$ bars, respectively. According to the pressure curve at supply and demand junctions in Case 2, the pressure at junction `a' (where the gas-fired unit is placed) reaches its lower bound during the period 5:45'-8:35' p.m. At the same time, the pressure at the supplier junction is also at its highest level. Since the pressure difference between these two junctions has reached its maximum, the fuel demand in Case 2 is not entirely met. On the other hand, it is noticed that in the same graph for Case 3, the pressure difference between supplier and demand junctions has not yet been constrained, and the natural gas network is able to deliver all of the requested gas demand this time.
\vspace{-0.35cm}
\subsection{\textbf{Case 4: Sensitivity Analysis of FRPs}}
\par
\begin{figure}[h!]
    \vspace{-0.55cm}
    \centering
    \includegraphics[width=0.95\linewidth]{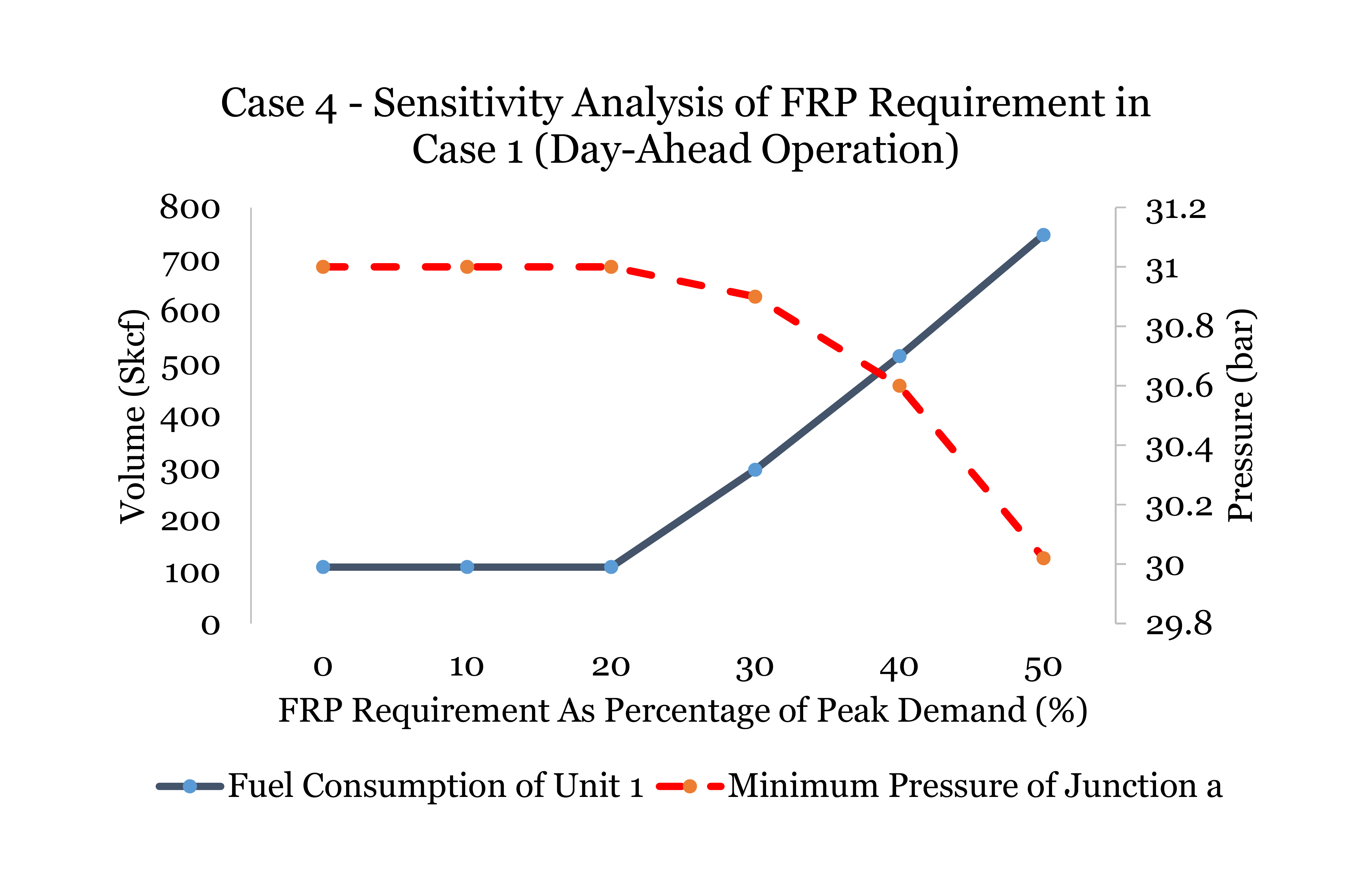}
    \vspace{-0.65cm}
    \caption{Effect of FRP requirement on fuel consumption and pressure}
    \label{fig:case4_fuel_pr}
    \vspace{-0.15cm}
\end{figure}

 This part illustrates the impact of different levels of FRP requirements on the coordinated day-ahead operation problem discussed in Case 1. The day-ahead operation results suggest that this network \textit{should be able to} provide FRP amounts equal to up to 50\% of the peak demand. The sensitivity analysis results for six different FRP requirement values are displayed in Fig. \ref{fig:case4_fuel_pr}. This plot displays the total daily fuel consumption of the gas-fired unit, along with the minimum daily pressure at junction `a' where the gas-fired unit is placed. As discussed in Case 2, when the pressure at the supplier junction reaches its upper bound, the pressure at the demand junction is an indicator of the system's ability to support the fuel demand. If the system-wide minimum and maximum pressures are at their bounds, the natural gas network is no longer physically capable of transferring additional gas.
\par It is noticed that at FRP requirement levels of 0\%, 10\%, and 20\%, the fuel consumption and minimum daily pressure of the system remain the same. The reason is that the non-gas-fired generator (unit 2) can provide the FRP requirements of the system on its own for these values. Here, the fuel consumption of the gas-fired unit is due to its generation during the peak period to balance the net demand. By increasing the FRP requirement from 20\% (6 MW per 5-minute interval) to 30\% (9 MW per 5-minute interval), since the non-gas-fired unit can offer only as much as 6.25 MW of FRP, the gas-fired unit has to come online to provide the remaining required FRP, hence the jump in the fuel consumption of this unit and the drop in minimum pressure at its junction. By increasing the FRP requirement from 30\%, the same pattern continues to occur. It is observed that with a 50\% FRP requirement, the minimum pressure drops to 30.02 bar. This suggests that the highest level of FRP that the natural gas network constraints allow for is 50\%. Nevertheless, it should be noted that the FRP model does not consider natural gas network constraints in the provision of the required fuel.

\par Another interesting observation in the sensitivity analysis of the FRP requirement is the amount of RES utilization throughout the day. It is noticed that higher FRP requirement levels lead to lower renewable utilization. Ideally, the operator would like zero renewable curtailments.
However, during periods when the output of RESs is high and electricity demand is low, curtailments occur. Table \ref{table:case4_renewable} shows the amount of renewable utilization for different FRP values. It is seen that with increasing the FRP requirement, the combined MWh output of generation units is rising, which subsequently results in lower renewable dispatch. The reason is that, with higher FRP requirements, the combined minimum output of both units must increase. Consequently, lower renewable output is required throughout all of the day. This phenomenon seems counter-intuitive; the FRP requirement is designed to reinforce renewable integration, whereas here it is noticed that higher FRP is achieved by decreasing renewable units' utilization.

\begin{table}[h!]
	\vspace{-0.35cm}
	\footnotesize \centering
	\caption {Renewable Utilization in Case 4} 
\begin{tabular}{ccccccc} \hline\hline
\textbf{FRP Requirement (\%)} & 0 & 10 & 20 & 30 & 40 & 50\\ \hline\hline
WT utilization (\%) & 95.9 & 95.7 & 95.4 & 95.0 & 94.7 & 94.4\\ \hline
PV utilization (\%) & 92.7 & 91.9 & 91.3 & 91.0 & 90.4 & 89.7\\ \hline
MWh All Units & 3462.6 & 3476.9 & 3491.4 & 3506.3 & 3521.6 & 3537.3\\ \hline
\hline
\label{table:case4_renewable}
\vspace{-0.75cm}
\end{tabular}
\end{table}

\subsection{\textbf{Case 5: Large Network Case}}
\begin{figure}[h!]
    \vspace{-0.35cm}
    \centering
     \includegraphics[width=1\linewidth]{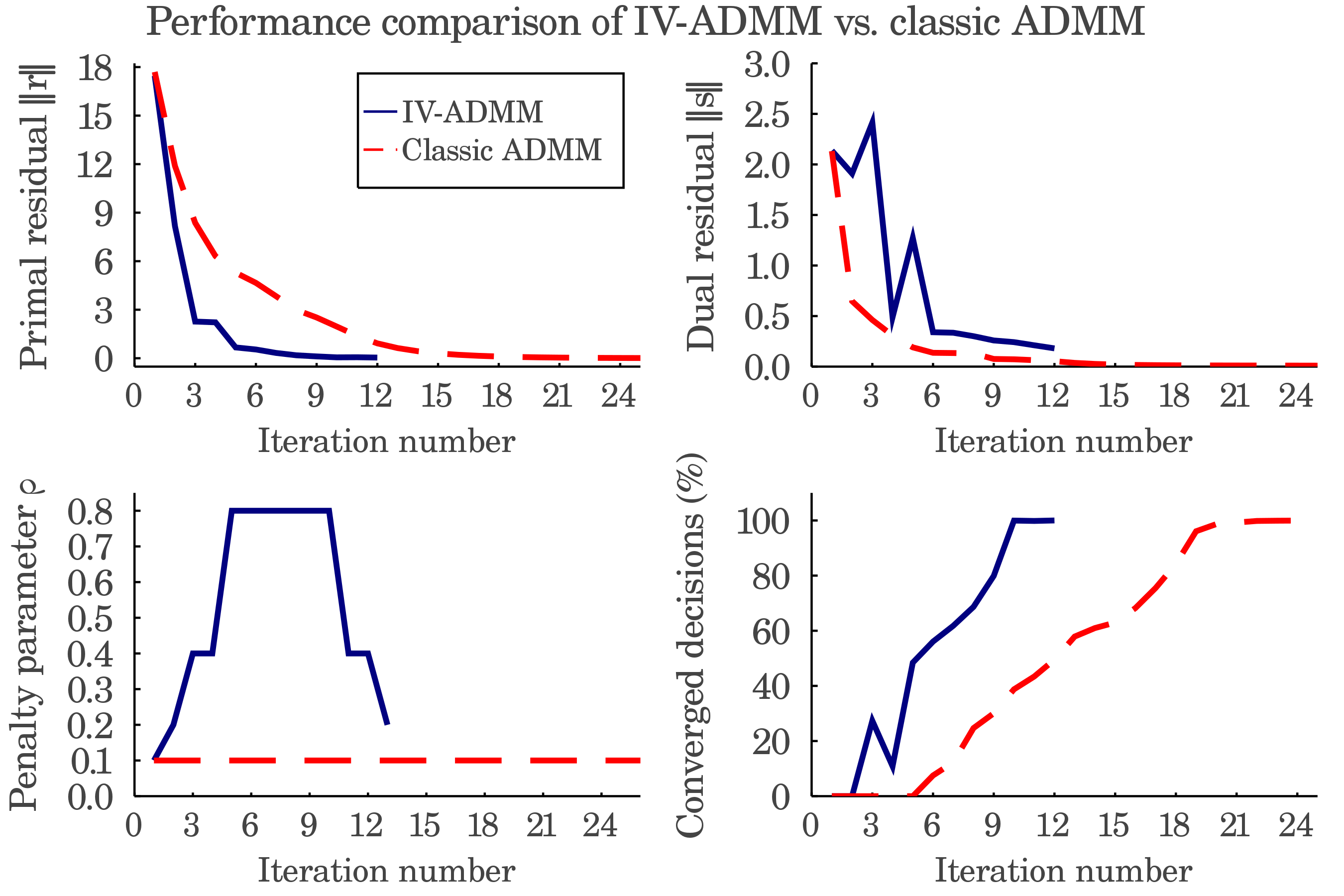}
    \vspace{-0.75cm}
    \caption{Performance comparison of IV-ADMM vs. classic ADMM}
    \label{fig:case5_admm_comparison}
  \vspace{-0.25cm}
\end{figure}

Finally, the coordinated day-ahead operation problem of a modified 118-bus transmission system integrated with a 10-junction natural gas network is considered to verify the proposed method's scalability. 
In this case, 8 out of 20 generation units are considered gas-fired. Here, the hourly FRP requirement is set at $25\%$ of the peak demand. 
It is revealed that all the electricity and natural gas demand are fully served in this system.
A comparison of the performance of IV-ADMM versus classic ADMM in Case 5 is displayed in Fig. \ref{fig:case5_admm_comparison}. According to these results, the classic ADMM converges in 25 iterations and 2725.9 seconds, while the presented IV-ADMM converges in 12 iterations and 789.6 seconds. This figure also displays how the value of $\left\|r\right\|$ is increased in the beginning stage to facilitate the reduction of the primal residual value and then decreased later on to stabilize the decisions. Here, all of the parameters are the same as the ones used in previous cases, with the exception of $\rho^0=0.1$, $\alpha=0.9$, and $\beta=2.8$.

\par
\begin{figure}[h!]
    \vspace{-0.35cm}
    \centering
    \includegraphics[width=0.9\linewidth]{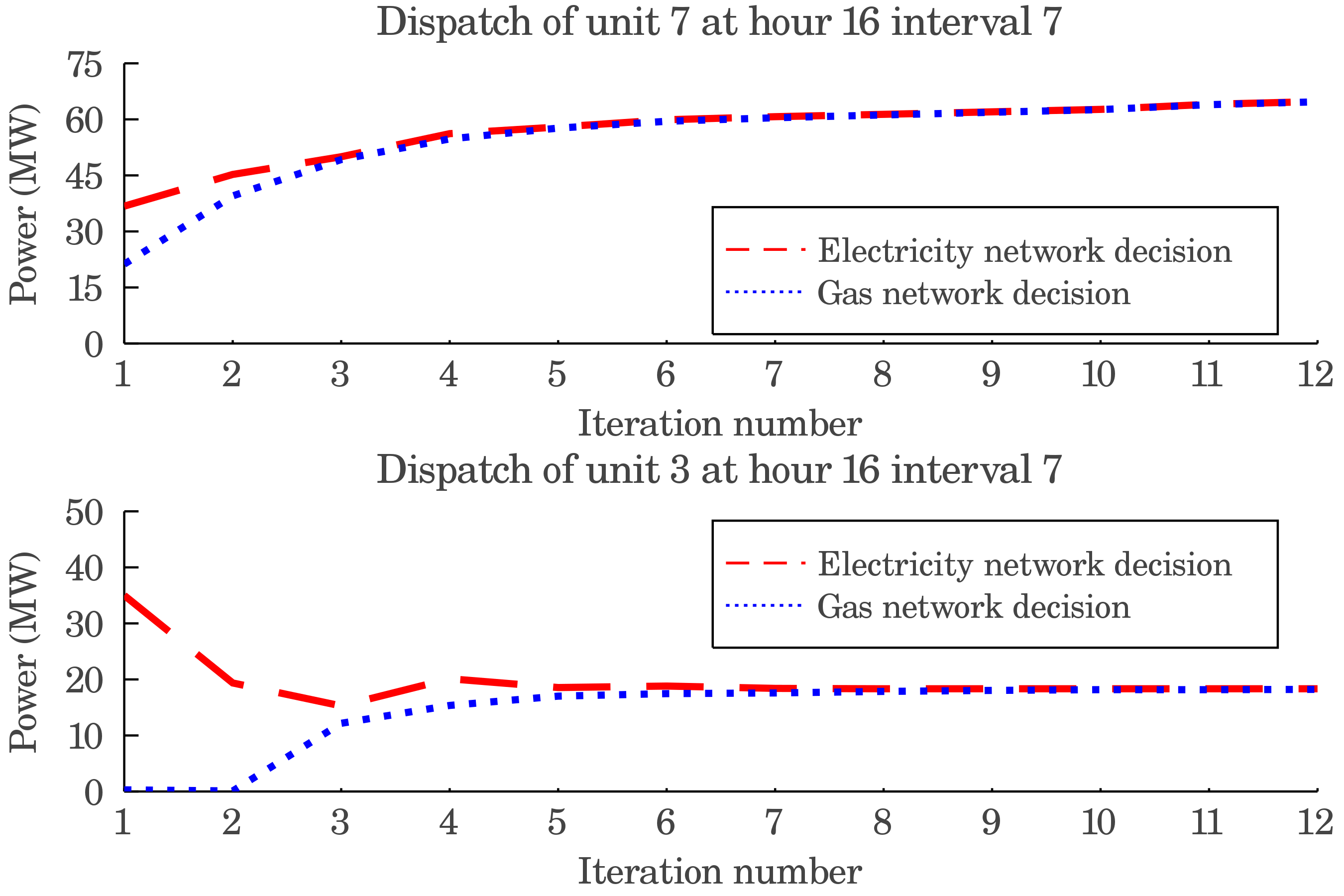}
    \vspace{-0.35cm}
    \caption{Decisions of Units 7 (top) and 3 (bottom) at a specific time}
    \label{fig:case6_converge}
    \vspace{-0.35cm}
\end{figure}

 Figure \ref{fig:case6_converge} shows the decisions for the output of units 7 and 3, passed between electricity and natural gas networks at each step of the IV-ADMM algorithm. Interestingly, it is observed that the consensus on the output of unit 7 is higher than both of the initial dispatches obtained by electricity and natural gas networks. Conversely, the consensus on the decision of unit 3 is lower than its initial dispatch. 
According to the initial dispatch obtained by the electricity network, the initial optimal dispatches of units 7 and 3 are 36.8 and 35.0 MW, respectively. However, the natural gas network is not capable of operating at this set-point. Consequently, a re-dispatch takes place gradually, which adjusts the decisions of all units to a new set-point where all of the electricity demand is still met, while the natural gas network constraints are satisfied and the required fuel is fully delivered. Here, the final decisions for the dispatch of units 7 and 3 are equal to 64.8 and 18.3 MW, respectively. 
\section{Conclusions}\label{sec:conc}
A new approach is presented for the coordinated operation of interconnected electricity and natural gas networks. The major advantages of the proposed method are utilizing a dynamic gas flow model, investigating the impact of FRPs, and applying a new distributed optimization approach. The results showed that our approach speeds up the convergence of decisions by more than three times compared to the classic method. To deal with the abrupt changes introduced by renewable generation integration, the flexible ramping product model is utilized. The cases showed that flexible ramping can prepare the network for variations of up to 50\% of the net demand. However, through a hypothetical scenario, it was shown that ramp products are not a reliable means of offering flexibility in interconnected electricity and natural gas networks. This is because the flexible ramp model neglects the governing dynamics of the natural gas network and takes fuel delivery to gas-fired units for granted. Cases are dedicated to the analysis of natural gas network dynamics. We illustrate how pressures at different locations of a natural gas network affect the amount of natural gas that is fed at demand junctions. In another observation, it was noticed that increasing the flexibility of the electricity network leads to higher renewable curtailments. The results of this article suggest that in order to reliably apply flexible ramping products to the coordinated operation of coupled electricity and natural gas networks, a new model should be developed which can account for the implications of ramping services for natural gas network dynamics.

\bibliographystyle{IEEEtran.bst}
\bibliography{ramp_dynamics/journal_ramp}

\end{document}